\documentclass{aa}  

\usepackage{graphicx}
\usepackage{txfonts}
\usepackage{xspace}
\newcommand{\flux}{$\rm erg\,\rm cm^{-2}\,\rm s^{-1}$\xspace}
\newcommand{\nicer}{\textit{NICER}\xspace}
\newcommand{\xmm}{\textit{XMM-Newton}\xspace}
\newcommand{\nustar}{\textit{NuSTAR}\xspace}
\newcommand{\swift}{\textit{Swift}\xspace}

\usepackage{xcolor}
\usepackage{tablefootnote}
\usepackage{subfigure}
\begin{document}

\title{Joint ALMA/X-ray monitoring of the radio-quiet type 1 AGN IC\,4329A}

\author{E. Shablovinskaya \thanks{\email{elena.shablovinskaia@mail.udp.cl}}\inst{1}
    \and C.~Ricci\inst{1,2}
    \and C-S.~Chang\inst{3}
    \and A.~Tortosa\inst{4}
    \and S.~del~Palacio\inst{5}
    \and T.~Kawamuro\inst{6}
    \and S.~Aalto\inst{7}
    \and Z.~Arzoumanian\inst{8}
    \and M.~Balokovic\inst{9}
    \and F.~E.~Bauer\inst{10,11,12}
    \and K.~C.~Gendreau\inst{13}
    \and L.~C.~Ho\inst{2,14}
    \and D.~Kakkad\inst{15}
    \and E.~Kara\inst{16}
    \and M.~J.~Koss\inst{17,12}
    \and T.~Liu\inst{18}
    \and M.~Loewenstein\inst{19,20}
    \and R.~Mushotzky\inst{20,21}
    \and S.~Paltani\inst{22}
    \and G.~C.~Privon\inst{23,24,25}
    \and K.~Smith\inst{26,27}
    \and F.~Tombesi\inst{28,4,29}
    \and B.~Trakhtenbrot\inst{30}
     }

   \institute{Instituto de Estudios Astrofísicos, Facultad de Ingeniería y Ciencias, Universidad Diego Portales, Santiago 8370191, Chile
    \and Kavli Institute for Astronomy and Astrophysics, Peking University, Beijing 100871, China
    \and Joint ALMA Observatory, Avenida Alonso de Cordova 3107, Vitacura, Santiago 7630355, Chile
    \and INAF -- Astronomical Observatory of Rome, Via Frascati 33, 00040 Monte Porzio Catone, Italy
    \and Department of Space, Earth and Environment, Chalmers University of Technology, SE-412 96 Gothenburg, Sweden
    \and RIKEN Cluster for Pioneering Research, 2-1 Hirosawa, Wako, Saitama 351-0198, Japan
    \and Department of Space, Earth and Environment, Chalmers University of Technology, Onsala Space Observatory, 43992 Onsala, Sweden
    \and X-Ray Astrophysics Laboratory, NASA Goddard Space Flight Center, Code 662, Greenbelt, MD 20771, USA
    \and Yale Center for Astronomy \& Astrophysics, 52 Hillhouse Avenue, New Haven, CT 06511, USA
    \and Instituto de Astrof{\'{\i}}sica and Centro de Astroingenier{\'{\i}}a, Facultad de F{\'{i}}sica, Pontificia Universidad Cat{\'{o}}lica de Chile, Campus San Joaquín, Av. Vicuña Mackenna 4860, Macul Santiago 7820436, Chile
    \and Millennium Institute of Astrophysics, Nuncio Monse{\~{n}}or S{\'{o}}tero Sanz 100, Of 104, Providencia, Santiago, Chile 
    \and Space Science Institute, 4750 Walnut Street, Suite 205, Boulder, Colorado 80301, USA
    \and NASA Goddard Space Flight Center, Greenbelt, MD 20771, USA
    \and Department of Astronomy, School of Physics, Peking University, Beijing 100871, China
    \and Space Telescope Science Institute, 3700 San Martin Drive, Baltimore, MD 21218, USA
    \and MIT Kavli Institute for Astrophysics and Space Research, Massachusetts Institute of Technology, Cambridge, MA 02139, USA
    \and Eureka Scientific, 2452 Delmer Street Suite 100, Oakland, CA 94602-3017, USA
    \and Department of Physics and Astronomy, West Virginia University, P.O. Box 6315, Morgantown, WV 26506, USA
    \and Center for Research and Exploration in Space Science and Technology, NASA/GSFC, Greenbelt, MD 20771, USA
    \and Department of Astronomy, University of Maryland, College Park, MD 20742, USA 
    \and Joint Space-Science Institute, University of Maryland, College Park, MD 20742, USA
    \and Department of Astronomy, University of Geneva, ch. d'Écogia 16, CH-1290, Versoix, Switzerland
    \and National Radio Astronomy Observatory, Charlottesville, VA 22903, USA
    \and Department of Astronomy, University of Florida, Gainesville, FL, 32611, USA
    \and Department of Astronomy, University of Virginia, Charlottesville, VA, 22904, USA
    \and Department of Physics and Astronomy, Texas A\&M University, College Station, TX 77845, USA
    \and Department of Physics, Southern Methodist University, 3215 Daniel Ave., Dallas, TX 75205, USA
    \and Physics Department, Tor Vergata University of Rome, Via della Ricerca Scientifica 1, 00133 Rome, Italy
    \and INFN -- Rome Tor Vergata, Via della Ricerca Scientifica 1, 00133 Rome, Italy 
    \and School of Physics and Astronomy, Tel Aviv University, Tel Aviv 69978, Israel
    }

   \date{Received ...; accepted ...}

 
  \abstract
  {
  The origin of a compact millimeter (mm, 100--250~GHz) emission in radio-quiet active galactic nuclei (RQ AGN) remains debated. Recent studies propose a connection with self-absorbed synchrotron emission from the accretion disk X-ray corona. We present the first joint ALMA ($\sim$100~GHz) and X-ray (\nicer/\xmm/\swift; 2--10\,keV) observations of the unobscured RQ AGN, IC~4329A ($z = 0.016$). 
  The time-averaged mm-to-X-ray flux ratio aligns with recently established trends for larger samples \citep{Kawamuro22,ricci23}, but with a tighter scatter ($\sim$0.1~dex) compared to previous studies. However, there is no significant correlation on timescales of less than 20 days. 
  The compact mm emission exhibits a spectral index of $-0.23 \pm 0.18$, remains unresolved with a 13 pc upper limit, and shows no jet signatures. Notably, the mm flux density varies significantly (factor of 3) within 4 days, exceeding the contemporaneous X-ray variability (37\% vs. 18\%) and showing the largest mm variations ever detected in RQ AGN over daily timescales. The high amplitude variability rules out scenarios of heated dust and thermal free--free emission, pointing toward a synchrotron origin for the mm radiation in a source of $\sim$1 light day size. While the exact source is not yet certain, an X-ray corona scenario emerges as the most plausible compared to a scaled-down jet or outflow-driven shocks.}


   \maketitle
%

\section{Introduction}\label{sec:intro}

Most active galactic nuclei (AGN) do not show bright radio structures associated with powerful relativistic jets launched from regions close to the accreting supermassive black hole (SMBH). 
However, these radio-quiet (RQ) sources exhibit faint radio emissions, often appearing in compact, parsec-scale regions that remain unresolved even with VLBA observations \citep[e.g.][]{alhosani22,Chen23,wang23}. While in radio-loud AGN the emission is known to be dominated by non-thermal synchrotron radiation from jets, the origin of compact radio emissions in RQ AGN is still a subject of debate. Various hypotheses have been proposed, including small-scale jets, nuclear star formation regions, thermal free--free radiation, and the magnetised corona of the accretion disk \citep[see][for a review]{panessa19}. Among these explanations, the accretion disk corona is one of the most favoured. This arises from the RQ AGN following the radio/X-ray relation $L_\mathrm{5GHz} / L_\mathrm{0.2-20keV} \sim 10^{-5}$ \citep{laor08} previously discovered for the coronally active stars \citep{guedel93}, where the corona is magnetically heated similarly to what is expected for AGN \citep[e.g.][]{merloni01a,merloni01b}. In this scenario, the population of high-energy electrons upscattering UV/optical seed photons into X-ray energies also produces optically thick synchrotron radio emission while moving in the coronal magnetic field.
Under certain assumptions, 
the synchrotron emission of the structure with an approximately X-ray corona size can be self-absorbed at frequencies lower than $\sim$100--300~GHz (i.e., millimeter-wave band, mm), where the contribution from cold dust emission is minimal. 
Recent theoretical calculations \citep[e.g.,][]{Raginski16,Inoue18} also predict that coronal emission can peak at 100--200~GHz, producing flat synchrotron emission up to 300~GHz. Flux densities at 100\,GHz have been measured for a few dozen AGN and consistently exceed extrapolations from low-frequency power-law slopes \citep[e.g.,][]{behar15,behar18,Inoue18}, confirming the presence of a compact, optically-thick core.

\begin{figure*}
    \centering
    \includegraphics[scale=0.6]{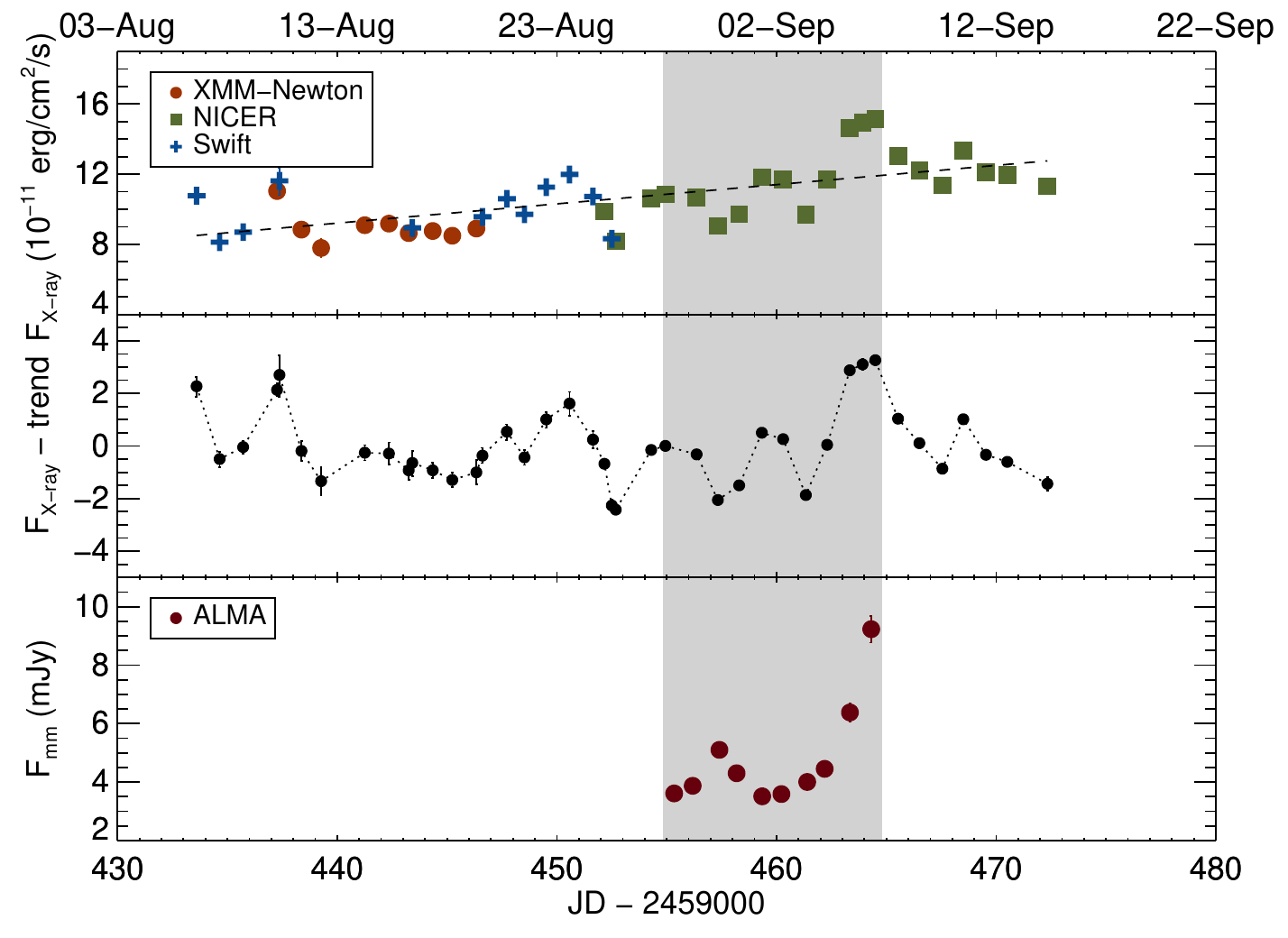}
    \caption{X-ray and mm-band light curves. From top to bottom: the X-ray (2--10~keV) flux with fitted power-law model, 2--10~keV flux deviations from the long-term trend, and the 3~mm (97.5~GHz) flux density. 
    The interval of the contemporaneous X-ray and mm observations is marked with a grey region. All data are listed in Appendix \ref{appendA}.}
    \label{fig:LC}
\end{figure*}

Recent research has further supported the connection between mm and X-ray luminosity: while \citet{behar18} found only a tentative correlation between 100~GHz and 2--10~keV luminosity for 34 AGN with a large scatter\footnote{Here and after we mean 1$\sigma$ scatter.} ($\sim$0.5~dex) likely caused by the sample heterogeneity and low angular resolution ($\gtrsim 1''$), a recent high-spatial resolution study of a sample of 98\,AGN with ALMA \citep{Kawamuro22,Kawamuro23} measured a highly significant linear correlation between the time-averaged hard X-ray (14--150~keV) and the instantaneous mm (230~GHz) luminosity with a $\sim$0.36~dex scatter. Notably, higher spatial resolution (1--23~pc) 100\,GHz ALMA observations of nearby hard X-ray-selected AGN have shown that the correlation between X-ray and mm emission \citep{ricci23} has a low scatter of 0.22~dex. However, to unequivocally establish the common origin of mm and X-ray flux by the electrons in the X-ray corona, it would be crucial to detect their correlated variability. 
Detecting correlated variability would also significantly contribute to our understanding of the X-ray origin, providing evidence that the magnetic field plays an important role in relativistic particle acceleration and corona heating \citep{merloni01a,merloni01b}.  Currently, only two observational campaigns \citep{behar20,petrucci23} have been conducted using IRAM (with $\sim$17--28$''$ resolution) and NOEMA ($\sim$1$''$) at 100~GHz, respectively, and found no definitive evidence of correlated variability (see Section \ref{lit}). In this context, sub-arcsecond observations with ALMA can play a crucial role due to its sensitivity and angular resolution to localise the compact source of the mm emission.

In this paper, we study the X-ray and mm variability of the nearby RQ AGN IC~4329A ($z = 0.016$) using, for the first time, ALMA observations with angular resolution $<$0\farcs1. Previous high-resolution (0\farcs16--0\farcs4) ALMA observations have shown that IC~4329A exhibits relatively bright ($\sim$8~mJy) compact emission at 100\,GHz, showing an excess compared to lower frequencies \citep{Inoue18}, which can be attributed to self-absorbed synchrotron radiation. Additionally, IC~4329A stands out as the brightest unobscured AGN in the southern sky \citep[$F_{2-10 \ \rm keV} \sim 10^{-10}$~\flux,][]{ricci17ApJS}, showing significant variability in the X-ray band \citep[from $\sim$8 $\times$ 10$^{-11}$~\flux to $\sim$2 $\times$ 10$^{-9}$ \flux in 2--10\,keV band on timescales as short as several hours, see][for details]{tortosa23}. All these characteristics, in both the X-ray and mm band, make IC~4329A the best target for the search for correlated X-ray/mm variability. In 2021, IC~4329A was the focus of an extensive observational campaign, with 45 X-ray observations carried out by \xmm, \nustar, \nicer and \swift and 10 ALMA observations over 10 consecutive days. The analysis of these high-quality \xmm and \nustar observations were described in \citet{tortosa23}. In this article, we report the analysis of the X-ray and mm variability on timescales of days, with the goal of investigating the correlation between these two bands. Throughout the paper we used the standard cosmological parameters ($H_0 = 70$~km\,s$^{-1}$\,Mpc$^{-1}$, $\Omega_{\rm m} = 0.3$, $\Omega_\Lambda = 0.7$).

\section{Observations and data reduction}

\subsection{X-ray}

\subsubsection{\xmm}
IC\,4329A was observed once per day for ten consecutive days (from 2021-08-10 to 2021-08-19, P.I. C. Ricci) by the X-ray Multi-Mirror Mission (\xmm, \citealt{Jansen2001}) during {\it XMM-Newton} AO\,19. The European Photon Imaging Camera (EPIC hereafter) cameras were operated in the small window and thin filter mode. The observation 0862090401 is not included in the analysis since, due to a problem in the ground segment, the EPIC exposure was lost.

The EPIC camera event lists are extracted with the \textsc{epproc} (pn) and and \textsc{emproc} (MOS) tools of the standard System Analysis Software (\textsc{SAS} v.18.0.0; \citealt{Gabriel2004}). The latest calibration files, available at January 2023, were used. The choice of optimal time cuts for the flaring particle background has been performed by visually inspecting light curves created in the energy range 10--12\,keV (EPIC-pn) with PATTERN=0 (single events). For the choice of source and background extraction radii, we identified point-like sources in each target field of view running the meta-task \texttt{edetect-chain} on the 0.5--2\,keV energy band EPIC images and performed an iterative process that maximises the signal-to-noise ratio (SNR) as in \citet{Piconcelli2004}. The resulting optimal source extraction radius was $30\arcsec$, and the background spectra were extracted from source-free circular regions with radii of $\sim$ $50\arcsec$ for each observation. Response matrices and auxiliary response files were generated using the SAS tools \textsc{rmfgen} and \textsc{arfgen}, respectively. EPIC-pn spectra were binned to oversample the instrumental energy resolution by a factor larger than three and to have no less than 20 counts in each background-subtracted spectral channel. No significant pile-up affected the EPIC data, as indicated by the SAS task \textsc{epatplot}.

\subsubsection{NICER}

Data reduction of the 21 NICER \citep{Gendreau12,Arzoumanian14,Gendreau16} observations (PI: C. Ricci) of our campaign was performed following the same procedure reported in \citet{Ricci21}.

\subsubsection{Swift}
A total of 12 observations from the X-ray telescope (XRT, \citealp{Burrows05}) on board the {\it Neil Gehrels Swift Observatory} \citep{Gehrels04} (PI: C. Ricci) were used here. We performed the \textit{Swift}/XRT data analysis using the \textsc{xrtpipeline} following the standard guidelines \citep{Evans09}.

\subsection{mm}

The ALMA band-3 observations were taken for 10 consecutive days from August 28 to September 6, 2021 (Project code 2019.1.01181.S; PI C. Ricci) in four channels of 1.985~GHz bandwidth with the central frequencies $\nu =$ 90.52, 92.42, 102.52, and 104.48~GHz. Most observations are taken within 3~hours from the \nicer observations. Data are processed using \texttt{CASA} version 6.1.1.15 and the ALMA Pipeline version 2020.1.0.40 \citep{alma_pipeline}. The spectral setup is in Time-Division Mode (TDM) around 100\,GHz, as we aim to detect the continuum emission of IC\,4329A. Our monitoring campaign was scheduled during the ALMA long baseline configuration (C43-9/10) with the longest baseline of 13.8\,km. For each observing session, the on-source time is about 8 minutes. As a result, the typical beam size is between 0$''$.04 and 0$''$.09 (i.e. $\sim$13 and $\sim$29 pc). According to the ALMA proposer’s guide, the statistical flux error for band-3 observations is 5\%. By analyzing the flux measurements of the phase calibrator J1351-2912, we derived consistent results. Therefore, we applied the 5\% error to our analysis. 

To have an insight into the spectral variability of IC~4329A, we derived the spectral index using the fluxes measured by the four spectral windows from the 10 epochs of observations. This analysis assumes that the spectral slope of the flux calibrator J1337-1257 and the phase calibrator were stable during our campaign. The ALMA calibration shows that the variation of the spectral index for the flux calibrator is within 2\% during our campaign.

\section{Results}

The X-ray (2--10~keV) and mm (100~GHz) light curves of IC~4329A obtained are illustrated in Fig.~\ref{fig:LC}. Note that the error bars are shown in the figure but are typically smaller than the symbols. In both wavebands, the observations were carried out with a daily cadence. The X-ray light curve, obtained with \textit{XMM-Newton}, \textit{NICER} and \textit{Swift}, spanning 40 days, is shown in the top panel of Fig.~\ref{fig:LC}. Contemporaneous ALMA 100\,GHz observations were carried out over 10 days (bottom panel). In the following, we examine the behavior of both light curves.

\begin{figure}
    \centering
     \includegraphics[scale=0.47]{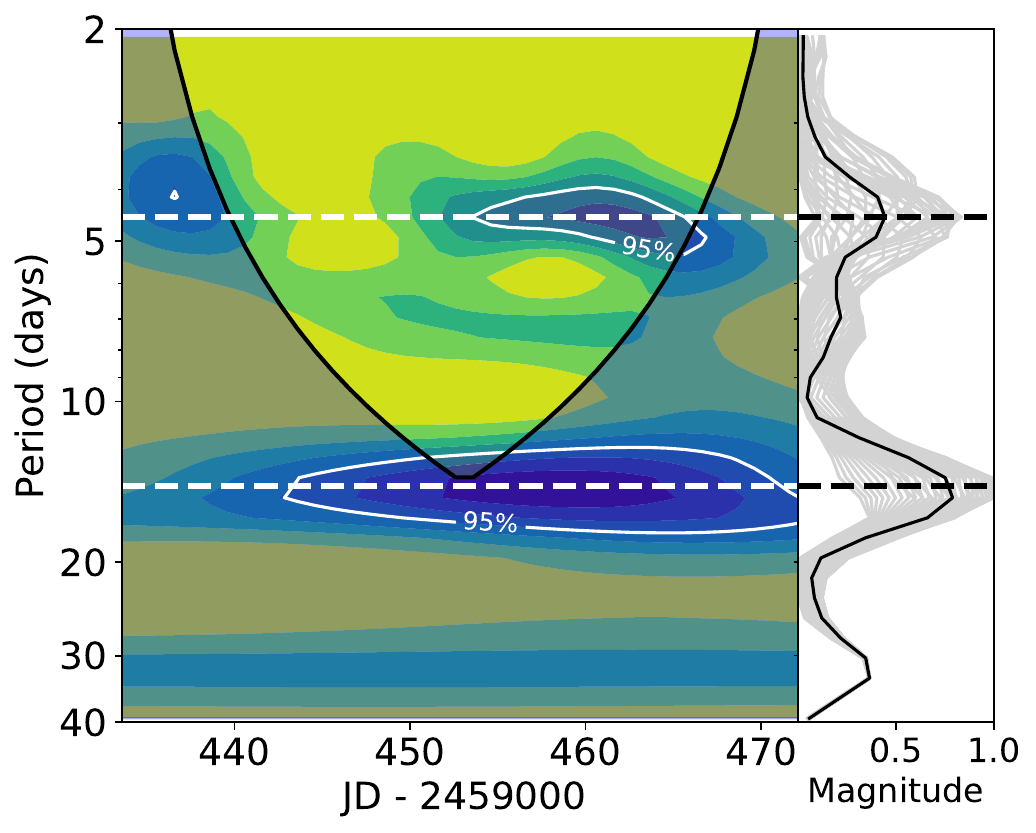}
    \caption{Wavelet analysis of the X-ray variability with the trend subtracted and the data interpolated on the regular grid. \textit{Left}: the magnitude of the wavelet transform. The 95\% confidence level is given in white contours. The cone-of-influence is given in the solid black line; the region out of the cone in the darker colour shows less reliable periods that may be affected by edge effects (see text for details). \textit{Right}: the wavelet transform profiles in grey and their median value in black indicate the variability periods of 14.9 and 4.5 days. The dashed lines correspond to the maxima of the Lomb-Scargle periodogram.}
    \label{fig:ICwav}
\end{figure}

\subsection{X-ray variability}

During our campaign, the X-ray (2--10~keV) flux of IC~4329A showed significant variability on timescales of days. To quantify the intensity of the variability taking into account the measurement uncertainties we calculated the fractional variability of the light curve $F_{\rm var}$ \citep{vaughan03}:
\begin{equation}
F_{\rm var}=\sqrt{\frac{S^2-\overline{\sigma^2}}{\overline{x_i}^2}} \times 100\%,
\label{eq:fvar}
\end{equation}
where $S^2$ is the sample variance, $\overline{\sigma^2}$ is the mean square error, and $\overline{x_i}^2$ is the square mean of the data. For the X-ray data, $F_{\rm var}^\mathrm{X-ray} = (17.5 \pm 0.5)$\%, while the intraday variations are less than 5\% \citep[see][]{tortosa23}. Due to the focus of the work on the variability of the longer (daily) timescale, we ignored the intraday variations in the analysis.  

During our monitoring campaign, the X-ray flux steadily increased (top panel of Fig.~\ref{fig:LC}), following a long-term trend of $\sim$10$^{-12}$~\flux per day. Once this trend has been subtracted (middle panel of Fig.\,\ref{fig:LC}), the X-ray light curve shows variations on a timescale of several days. Although the nature of variability in AGN is stochastic, the observed fluctuations in the X-ray flux relative to the average may suggest the presence of regular oscillations. To investigate the periodicity of these variations, we calculated the Lomb-Scargle periodogram \citep{lomb, scargle} used for the analysis of unevenly sampled data employing a least-squares fitting procedure. The periodogram revealed two possible variability periods of {$14.4 \pm 0.9$ and $4.5 \pm 0.1$ days}. However, when the light curve is irregular and evolves over time, wavelet analysis \citep{Morlet1983} is more efficient than the periodogram. Unlike the Fourier transform, whose kernel is not time-localised, wavelet analysis dissects the data series into distinct frequency components and examines each segment with corresponding temporal resolution. Therefore, the wavelet analysis provides both variability frequency (or period) and its evolution during the observations. For IC~4329A, the result of the wavelet transformation of the X-ray flux variability with the subtracted trend is illustrated in Fig.~\ref{fig:ICwav}. Here, the original epochs are interpolated to the regular grid using the spline extrapolation. The wavelet analysis revealed two variation periods corresponding to the periodogram result: $4.5 \pm 0.8$ and $14.9 \pm 2.5$ days. The $14.9 \pm 2.5$ days period is stable during the observations but is greater than 1/3 of the monitoring period and is located out of the cone-of-influence \citep[see details in][]{wavelet_prac}, i.e., in the area where the analysis may be affected by data limitations (such as the finite duration of observations) and edge effects, thus it may be an artefact. The $4.5 \pm 0.8$ days period is less stable but can potentially reveal quasi-periodic oscillation in the AGN. Nevertheless, the estimated periods characterize only the observed segment of the light curve, the overall behaviour of which can be entirely stochastic.


\begin{figure*}
    \centering
    \includegraphics[scale=0.6]{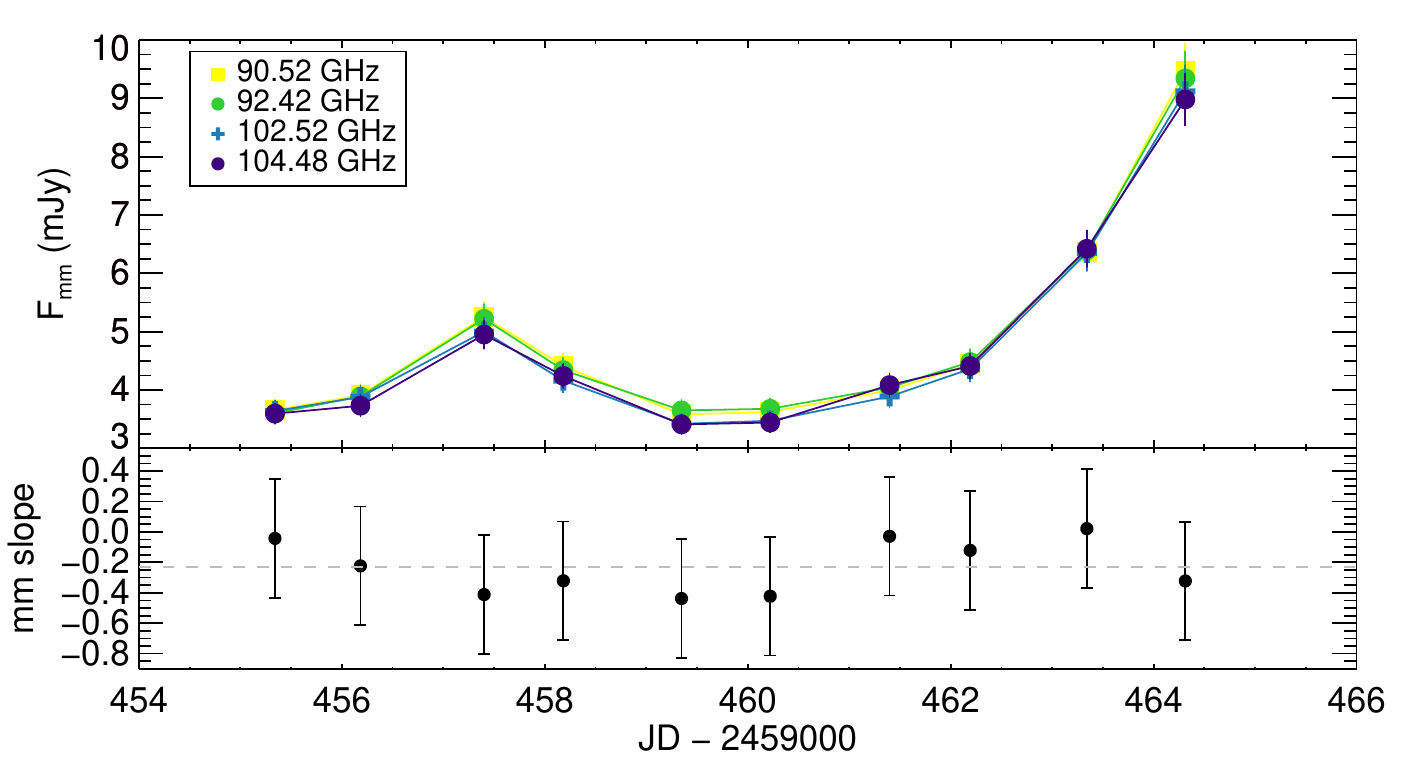}
    \caption{Upper panel: mm light curve in four different frequencies: 90.52, 92.42, 102.52, and 104.48~GHz, shown in different colours and symbols. Bottom panel: variations of the mm spectral slope. Grey dashed line corresponds to the median spectral slope $-0.23$.}
    \label{fig:ALMA_freq}
\end{figure*}

\subsection{Millimeter variability} \label{mm}

In the mm band, IC~4329A showed very strong variability during the 10-day observational campaign: the flux changed significantly, up to a factor of 2.6 between maximum and minimum, with the day-to-day variations reaching a factor of $\sim$1.5. The fractional variability for the mm data calculated using Eq.~(\ref{eq:fvar}) is two times higher than in the X-rays: $F_{\rm var}^{\rm mm} = (37.1 \pm 0.5)$\% including ALMA calibration uncertainties of $\sim$5\%. Even scanning the X-ray light curve with a 10-day window for a conservative comparison, $F_{\rm var}^{\rm X-ray}$ was found to be significantly lower than in the mm band, from 6.5\% to 18.5\%, and the maximum $F_{\rm var}^{\rm X-ray}$ was observed during the 10 days of the simultaneous X-ray/ALMA observations. Within 10 days, the mm light curve showed two clear maxima separated by $>$8 days. 

As ALMA observations were made at frequencies $\nu =$ 90.52, 92.42, 102.52, and 104.48~GHz, this could provide us with insight into the spectral variability of IC~4329A. Assuming a power-law spectrum ($F_{\rm mm} \propto \nu^\alpha$), for each epoch, we calculated the spectral slope $\alpha$. We found an average spectral slope $-0.23$ with $\sigma=0.18$, with time variations below $\sigma$. 

The light curves of IC~4329A in different frequencies are shown in Fig.\,\ref{fig:ALMA_freq} (upper panel). The variability pattern is the same for all wavebands. For comparison, we illustrated the variations of the spectral slope $\alpha$ in the bottom panel, Fig.\,\ref{fig:ALMA_freq} that did not show a correlation with the flux. Additionally, we examined the difference $\Delta$ between the lower and higher frequencies: $\Delta = (F_{91} - F_{103})/F_{103} \times 100$\%, where $F_{91}$ is the mean flux between 90.52 and 92.42~GHz, and $F_{103}$ is the mean flux between 102.52 and 104.48~GHz. 
Variations in this difference in time are statistically significant at the 95\% significance level, changing between $\sim$6\% (during the first flare) and $\sim$0\% (before the second flare). However, this value is also not correlated with the flux. 

\begin{figure}
    \centering
\includegraphics[width=\linewidth]{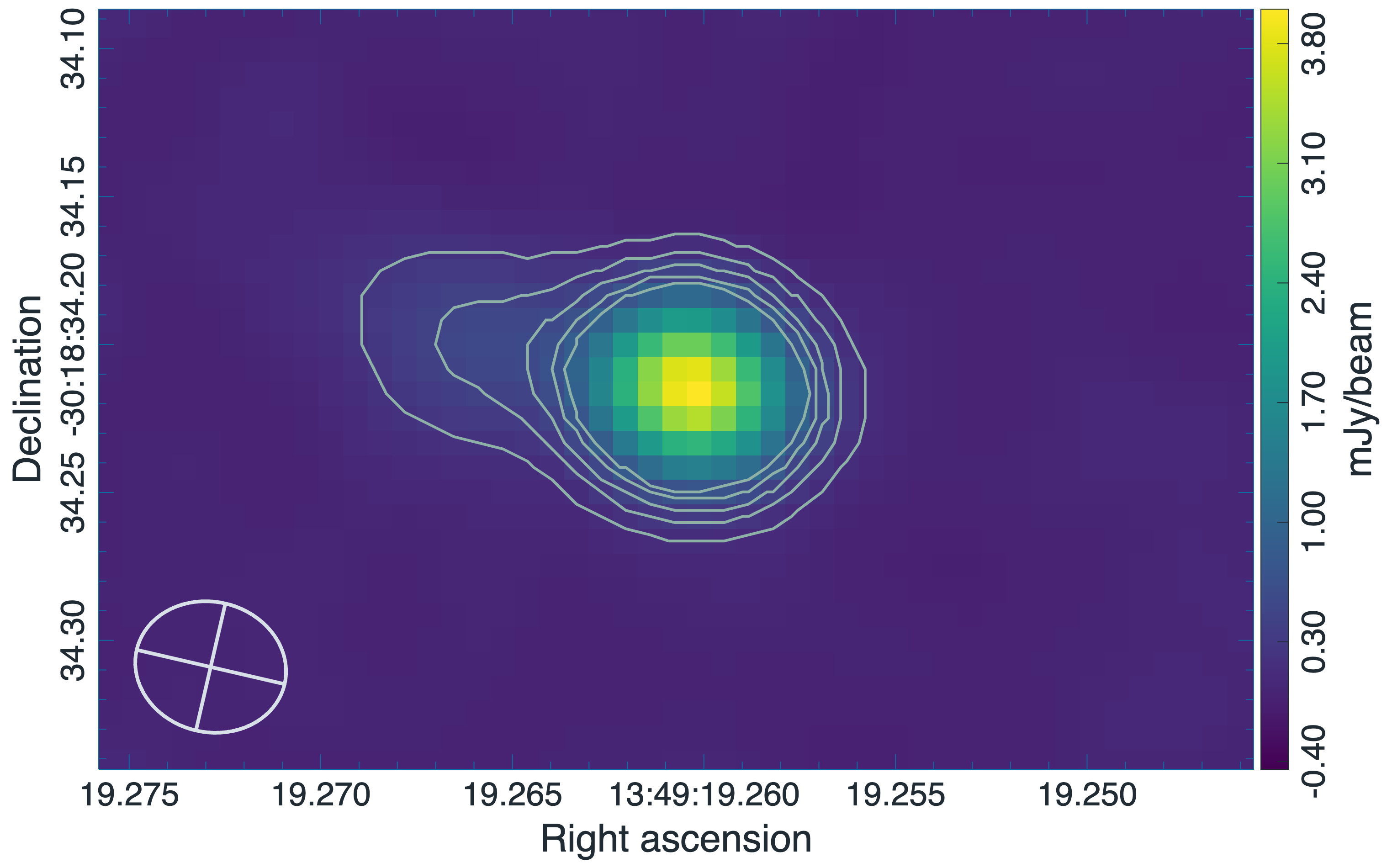}
    \caption{ALMA image of IC 4329A (median band 3) 9/3/2021. Beam size is 0\farcs05 $\times$ 0\farcs04 (i.e. $\sim$16 $\times$ 13 pc). The coordinates are given for the J2000 epoch.}
    \label{fig:map}
\end{figure}

The exceptional resolution and sensitivity of ALMA allowed us to resolve a faint diffuse structure to the east of the central bright source in the continuum maps of IC~4329A (Fig.~\ref{fig:map}). Notably, this structure is observed throughout the entire campaign, though due to beam size and shape, it is clearly distinguishable in only five of the 10 epochs. We measured the flux of this diffuse spot whenever possible and found that it is consistent, within the uncertainties in all epochs, with a flux density of $\sim$0.6~mJy. Given that this value is 10 times less than the flux emitted by the bright variable core even at its minimum, we ignored the contribution of this constant fainter elongated structure to the mm flux.

\subsection{X-ray -- mm correlation} \label{cor}

\begin{figure}
    \centering
     \includegraphics[scale=0.35]{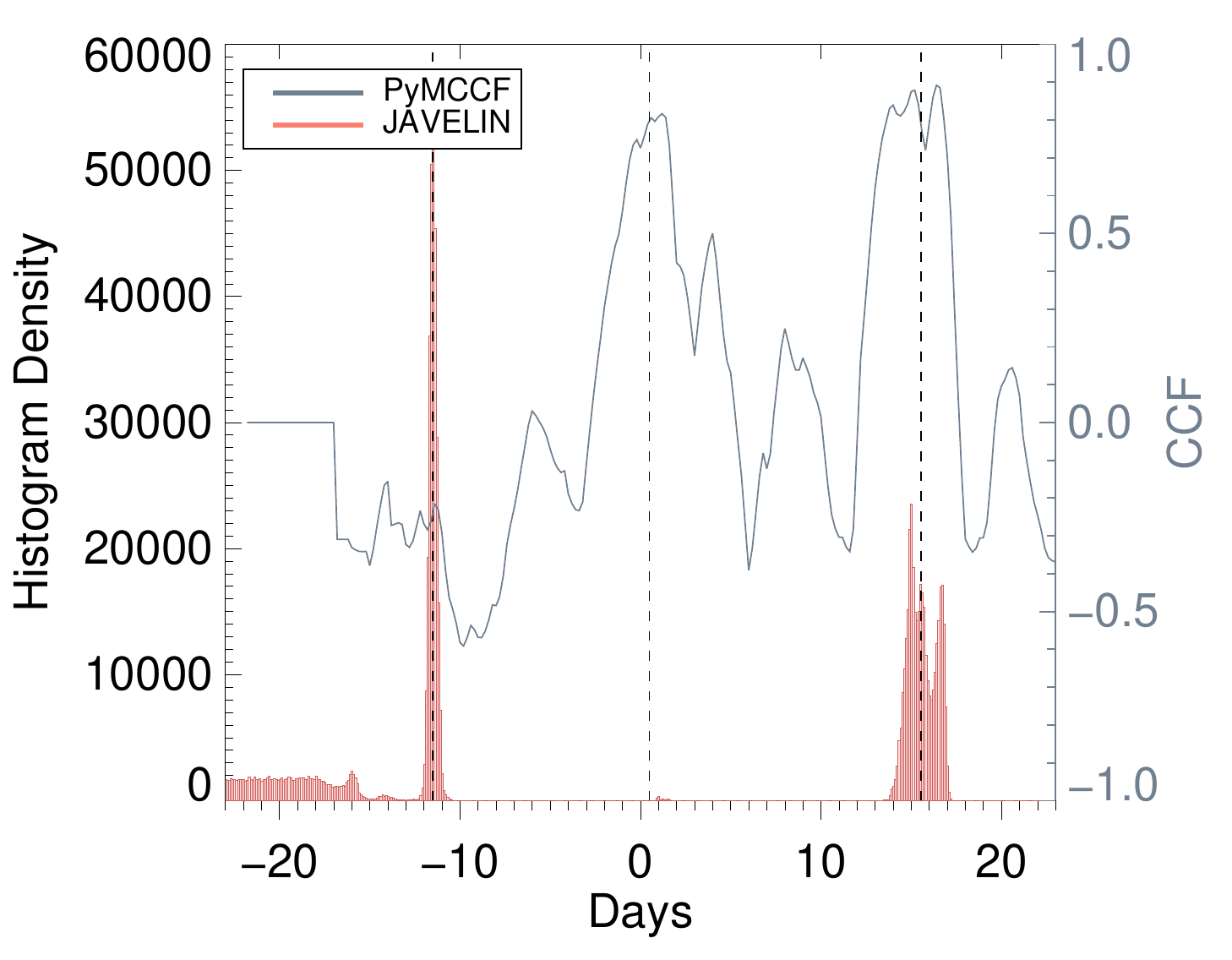}
    \caption{Results of the cross-correlation analysis of X-ray and mm data. The posterior distribution of the time lag provided by \texttt{JAVELIN} is illustrated by the red histogram. Two peaks at $\tau_1 = -11.5$ days (for the mm flux preceding the X-ray one) and $\tau_2 = 15.5$ days are marked with vertical dashed lines. The result of \texttt{PyMCCF} is shown in grey with the 0.5-day time lag marked with a vertical dashed line. }
    \label{fig:javIC}
\end{figure}

To quantitatively investigate a possible correlation and time delay between the X-ray and mm light curves, we applied several approaches commonly used in reverberation mapping campaigns \citep[as, e.g., SDSS-RM project, see][]{sdss-rm}. Here and in the rest of the paper, positive time lags correspond to the X-ray flux variability preceding the mm one. For the analysis, we used the full X-ray and mm light curves without trend subtraction. One of the most commonly used approaches to look for correlated variability is the interpolated cross-correlation function (CCF), which involves the linear interpolation of unevenly sampled light curves. Its modernised version \texttt{PyMCCF} \citep[][]{gaskell86,okn93,peterson98,pymccf} optimizes interpolation to reduce noise occurring from interpolation errors minimising the number of interpolated points used. The result of the 
\texttt{PyMCCF} analysis is shown as a grey curve in Fig.~\ref{fig:javIC}. The calculated CCF shows two peaks corresponding to time delays of $15.1^{+1.4}_{-0.7}$ and of $0.5^{+0.1}_{-0.1}$ days (vertical dashed lines). The errors are estimated by calculating 84.13th (+1$\sigma$) and 15.87th ($-$1$\sigma$) percentiles of a dataset. 

It has been shown that CCF-based methods can be less stable and efficient when the observational cadence is irregular \citep{Li19}. This is the reason why tools such as, e.g., \texttt{JAVELIN} \citep{zu16,yu20} have been widely used. 
\texttt{JAVELIN} constructs a model of the flux variability utilizing the damped random walk (DRW), extracting posterior distributions for critical DRW parameters through MCMC sampling. Another iteration of MCMC allows \texttt{JAVELIN} to determine the posterior time delay distribution between the light curves comparing the constructed models. Applying the \texttt{JAVELIN} analysis to the IC~4329A data, we found two peaks in the posterior distribution of the time lag (see the histogram of $10^5$ sampling iterations in Fig.~\ref{fig:javIC}), which are not identical to those identified by \texttt{PyMCCF}. We found two possible cases: a negative time lag $\tau_1 = -11.5 \pm$0.2 days and a positive time lag $\tau_2 = 15.5 \pm$0.9 days. The posterior distribution shows a larger and narrower peak at $-11.5$ days; however, the mm light curve shifted to $-11.5$ days does not fully overlap the X-ray one, so, $\tau_1$ comes partially from the modelled variability behaviour. Also, \texttt{JAVELIN} analysis generally provides one with a constraint on the width of a top-hat function. This quantity could be used to have an insight into the size of the structure responding to variable emission. However, for our dataset, we found no peak in the distribution of this parameter.

We also used the \texttt{PyROA} tool \citep{pyroa} based on a similar approach as \texttt{JAVELIN}, with the main difference being that in this case the variability is described using a running optimal average. 
However, it appeared that for the given data series the \texttt{PyROA} code provides unstable results, dependent on the initial parameters of the model, and no clear estimation of a time lag can be obtained. 

Generalizing the results obtained by various methods, we obtain three possible time delays: -11.5, 0.5, and 15.5. Note that 15 and 11 days are longer than the duration of the ALMA campaign, and 0.5 days is less than the cadence of the observations. Thus, the data is distorted by windowing and aliasing effects. While we will continue to analyze the light curves considering all three time lags, their reliability is questionable. Therefore, we will prioritize focusing on $\tau=0$.

\begin{figure*}
    \centering
    \includegraphics[scale=0.7]{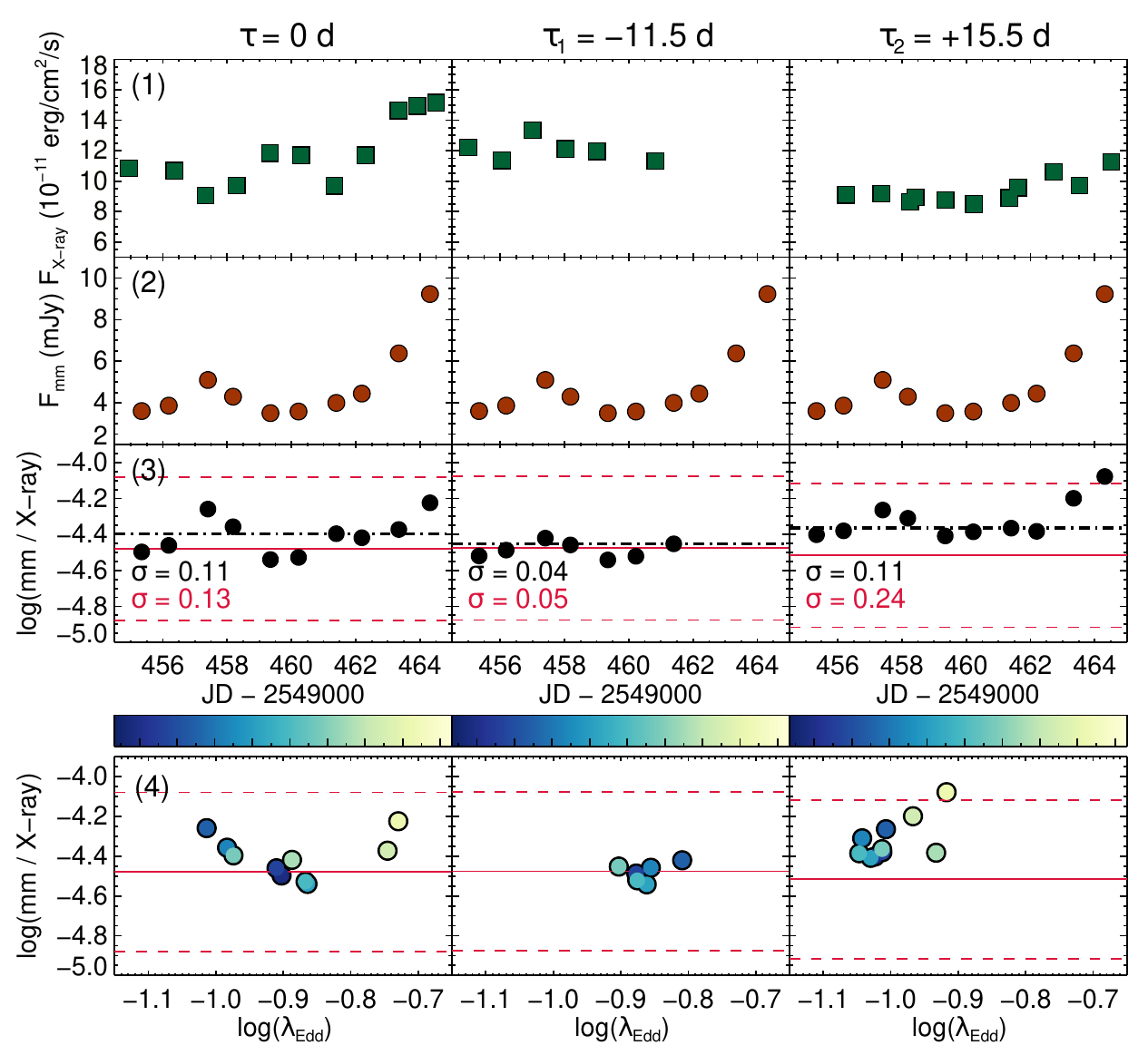}
    \caption{Variations of the mm/X-ray ratio during our observational campaign. The X-ray light curves shifted to the three time-lags: $\tau = 0$, $-11.5$ and +15.5 days (see columns from left to right) are shown in panel (1), and the mm light curve is in panel (2). The ratio $\log(F_{\rm mm}/F_{\rm X-ray})$, where the X-ray light curve was interpolated to the ALMA epochs, is given in panel (3) for each of the time-lags; the black dash-dotted line is the median value, and the red solid line corresponds to the relation between 100\,GHz and 2--10\,keV fluxes from \citep{ricci23} with 1$\sigma$ scatter given by the dashed red lines. The IC~4329A epoch-to-epoch scatter ($\sigma$) is also presented in the figure: the scatter relative to the median is shown in black, while the one relative to the predicted value is in red.  Panel (4) contains the same values as panel (3) but as a function of the Eddington ratio $\lambda_{\rm Edd}$. The colour of the circles illustrates the date of the observation. }
    \label{fig:evolution}
\end{figure*}



\section{Discussion}

\subsection{Evolution of mm/X-ray ratio \& intrinsic scatter} \label{evo}

Recently, for a volume-limited sample of radio-quiet hard X-ray selected AGN, \citet{ricci23} found a tight correlation between the observed X-ray (2--10\,keV) and mm (100\,GHz) emission, described by 
\begin{equation} \label{r}
\log F_{\rm mm} = (-0.8\pm0.4) + (1.37\pm0.04)\log F_{\rm X-ray},
\end{equation}
where the fluxes are given in units of erg s$^{-1}$ cm$^{-2}$, with a scatter of 0.22\,dex. From this correlation, the typical ratio between the 100~GHz continuum and the 2--10~keV emission is $\log(F_{\rm mm}/F_{\rm X-ray}) = -4.63 \pm 0.06$. 
Also, a significant correlation between 230~GHz and 14--150 keV luminosities for 98 AGN was found by \citet{Kawamuro22} with a scatter of $\sim$0.36\,dex. In both cases, non-simultaneous X-ray and mm observations were used, thus the scatter observed in these correlations can be (at least partly) due to the intrinsic variability of the AGN. 
To investigate intrinsic scatter in the correlation, we studied the variations of the mm to X-ray ratio in our IC~4329A data given in the left column on panels 1 and 2 of Fig.~\ref{fig:evolution}. Note that while the original X-ray and mm data were obtained with a minor time difference of a few hours, sometimes the data points cannot be compared with each other unambiguously. Therefore, we interpolated the X-ray data to match the ALMA epochs. 
The evolution of $\log(F_{\rm mm}/F_{\rm X-ray})$ over time during the simultaneous mm and X-ray monitoring is shown in Fig.\,\ref{fig:evolution} (panel 3, left column). 
The scatter was estimated relative to two average values: the first is the one expected from the Eq.~(\ref{r}) calculated for the mean X-ray flux $\log(F_{\rm mm}/F_{\rm X-ray}) = -4.5$, and the second is the time-averaged value of the flux ratio was found to be $\log(F_{\rm mm}/F_{\rm X-ray}) = -4.4$. The average ratio found is in good agreement with the X-ray/100~GHz correlation. We estimated the scatter to be 0.13~dex and 0.11~dex, respectively. 
In both cases, the scatter is about two times smaller than one from \cite{ricci23}. This suggests that non-simultaneous observations can introduce significant scatter in the mm/X-ray correlation due to the intrinsic variability of AGN, as discussed in \cite{Kawamuro22}.

Since we have found a possible time delay between the mm and X-ray light curves, in Fig. \ref{fig:evolution} we show the mm/X-ray ratio both for the zero time lag and two more cases: $\tau = -11.5$ and 15.5 days (middle and right column, respectively).
In panel 1, the X-ray light curve is shifted according to the different $\tau$, and the unshifted mm light curve is in panel 2.
We found that the scatter of the mm/X-ray ratio (panel 3) changes when we correct the light curves for the possible time shift. For $\tau = +15.5$, the scatter relative to the median ratio $\log(F_{\rm mm}/F_{\rm X-ray}) = -4.3$ is the same as for $\tau = 0$, but twice larger, 0.24~dex, relative to the value calculated with Eq.~(\ref{r}). For the $\tau = -11.5$ days, the scatter becomes very small ($\sigma = 0.05$), and the median ratio coincides with that predicted by the correlation. The values of scatter calculated as the standard deviation relative to the mm/X-ray average and the value predicted from mm/X-ray correlation from \cite{ricci23} are illustrated in Fig.\,\ref{fig:evolution} for each time lag.


Together with the evolution of the mm/X-ray ratio in time, we also studied its variation with the Eddington ratio ($\lambda_{\rm Edd}$). To calculate $\lambda_{\rm Edd}$, we used the SMBH mass recently estimated for IC\,4329A by reverberation mapping  $M_{\rm BH} = 6.8^{+1.2}_{-1.1} \times 10^7 M_{\odot}$ \citep{bentz4329}, while the bolometric luminosity was calculated from the X-ray flux in the 2--10\,keV range. We considered a comoving distance $D = 69.3$\,Mpc \citep{koss_catalogue}. Recently, it has been shown that the 2--10\,keV bolometric correction ($\kappa_{2-10}$) is primarily a function of the Eddington ratio (Gupta et al. 2024, in prep.), and it follows:
$$
\log(\kappa_{2-10}) = C \times \log(\lambda_{\rm Edd})^2 + B \times \log(\lambda_{\rm Edd}) + A,
$$
where $C = 0.054 \pm 0.034$, $B = 0.309 \pm 0.095$ and $A = 1.538 \pm 0.063$. Using this approach we found that during joint ALMA/X-ray monitoring, the Eddington ratio changes from $\log(\lambda_{\rm Edd}) = -1.01$ to $-0.73$ with a median value $\log(\lambda_{\rm Edd}) = -0.89$ for $\tau = 0$. The median bolometric correction is $\kappa_{2-10}=20.2$. The $\kappa_{2-10}$ could be tentative due to the large intrinsic dispersion, the obtained value of the bolometric correction is very close to the commonly used value $\kappa = 20$ from \cite{VF09}. 
In panel 4 of Fig. \ref{fig:evolution} we illustrate the relation between the mm/X-ray ratio and the Eddington ratio. The figure does not show a clear correlation between the mm/X-ray ratio and $\lambda_{\rm Edd}$ for the three different time lags. 

It is important to note that while our calculated time delays show lower scatters compared to previous studies, we have not observed a clear correlation between X-ray and mm emission over the campaign duration, which is not surprising given the short duration of ALMA observations and coarse sampling. This suggests that unstable processes might dominate on the scale of days, leading to varied flux behaviours. The discovered tentative characteristic timescales selected by the periodogram and wavelet analysis also do not shed light on this correlation: the observed mm flares are spaced 8 days apart, while the most probable X-ray oscillations are of the 5 days. Despite the roughly constant flux ratio between mm and X-rays, it seems different processes influence their variability on short timescales. With longer ALMA observations, we could better understand this flux correlation.

\subsection{Comparison with other mm/X-ray monitoring campaigns} \label{lit}

Simultaneous mm and X-ray studies of the variability of RQ AGN are still scarce, and only two such monitoring campaigns have been carried out before. The first of these was dedicated to the RQ Seyfert\,1 NGC\,7469 located at the same distance as IC~4329A ($z=0.016$) and having lower mass $M_{\rm SMBH} \approx 9.1 \times 10^6 M_\odot$ \citep{Peterson7469}. This object was known for a significant variability at 95~GHz, by a factor of two within 4--5 days as found in CARMA (Combined Array for Research in Millimeter-wave Astronomy) observations by \citet{Baldi15}. Later, \citet{behar20} provided the simultaneous mm/X-ray monitoring of NGC~7469. The observations were taken in the 0.3--10\,keV range with \textit{Swift}/XRT and in the mm band at 95 and 143~GHz frequencies with the Institut de Radioastronomie Millimetrique (IRAM) 30-m single-dish radio telescope. This monitoring lasted longer than our campaign ($\sim$70 days in X-rays, of which $\sim$50 days of contemporaneous mm and X-ray observations) with a cadence of 1--2 days, but both the typical mm uncertainties ($\sim$0.8~mJy) and beam sizes ($\sim$17--28$''$) are considerably larger than those of our ALMA observations. Yet \citet{behar20} stated a marginal correspondence of the measured mm flux with earlier observations with a smaller beam size $\sim$2.2$''$ at CARMA interferometer \citep{Baldi15}, the mm/X-ray ratio observed for NGC~7469 is unexpectedly large: $\log(F_{\rm 95 GHz}/F_{\rm 2-10 keV}) \approx -3.5$, which could be, at least in part, explained by the low resolution of both IRAM and CARMA data. The scatter associated with variability ($\sigma = 0.12$) appears consistent with that found for the current dataset from IC~4329A.  

While \citet{behar20} provides a comprehensive analysis of the X-ray and mm variability in NGC~7469, we are reanalyzing this data using the same approaches applied to our IC~4329A data. This ensures a homogeneous analysis of both datasets for a more accurate comparison. Using the wavelet analysis, we do not find any evidence of a periodicity in the X-ray light curve of NGC~7469. Concerning the time delay between X-ray and mm emission in the simultaneous NGC~7469 observations, \citet{behar20} cautiously stated that the mm variability appears to precede the X-ray flux by $\sim$14 days. We have repeated the correlation analysis of NGC~7469 data using the same tools as described in Sect.~\ref{cor}. The \texttt{PyROA} and \texttt{PyMCCF} methods did not reveal any time lag, while
the analysis done with the \texttt{JAVELIN} code confirmed the $-14$ days time lag. Moreover, we also discovered two more peaks in the posterior distribution, corresponding to 4.1 and 13.8 days. The correlation peaks at $-14$ and 13.8 days showed an amplitude approximately 7 times lower than the peak at 4.1 days. Thus, as in the case of IC~4329A, we cannot clearly determine the presence and magnitude of the time delay between mm and X-rays. While in NGC~7469, this uncertainty could result from insufficient observational accuracy, a similar outcome for IC~4329A indicates that in both cases, the relation between the bands is intricate, and shorter timescales of the variability need to be probed.

\citet{behar20} also examined the X-ray hardness ratios as a proxy of possible changes in the level of photo-electric absorption, which could give rise to the observed X-ray variability. However, neither the "harder when dimmer" trend, predicted by \citet{behar20} following \cite{Mehdipour17}, nor a significant correlation between mm flux and hardness ratio was observed in NGC~7469. Following \citet{behar20}, we calculated the hardness ratio in terms of count rates for the \textit{NICER} data for IC~4329A: HR = $(H - S)/(H + S)$, where $H$ is the count rate (count s$^{-1}$) in the hard (2.0--10.0~keV) band, and $S$ is the count rate in the soft (0.3--2.0~keV) band. We found that the HR does not vary significantly, and it does not show any correlation with the mm flux or mm spectral index changes.  

A second simultaneous mm/X-ray campaign of an RQ AGN was recently reported by \citet{petrucci23} for MCG+08$-$11$-$11 ($z=0.02$), a Seyfert~1 galaxy with a black hole mass comparable to that in IC~4329A, $M_{\rm SMBH} \approx 2.8\times 10^7 M_\odot$. The variability in the 3--10\,keV X-ray band obtained by \xmm\, and the 100\,GHz flux inferred by NOEMA were studied on time scales of 14\,hours. Both fluxes showed slight increases, corresponding to $\sim$1.06 times and $\sim$1.19 times for the mm and X-ray flux, respectively, but no correlated variability was found. Similarly to NGC\,7469 data, the mm/X-ray ratio is significantly larger than the mean value found for nearby AGN by \citet{ricci23}: $\log(F_{\rm 100 GHz}/F_{\rm 2-10 keV}) \approx -3.4$ with $\sigma = 0.12$.  Moreover, the mm/X-ray ratio previously measured in \citet{behar18} with comparable angular resolution in mm also appeared an order of magnitude larger than \citet{ricci23} predicted, yet the mm flux was 2.4 times lower ($\sim$7.5~mJy instead of $\sim$18.3~mJy). It is likely that such a large mm/X-ray ratio is due to the relatively low resolution of the mm observations ($\sim$1$''$ according to the data archive), for which other components emitting in mm may contribute to the measured flux. Therefore, the observed mm variability might be more contaminated by non-variable extended emission. This highlights the critical importance of utilizing ALMA, not only for achieving higher accuracy of the flux measurement (twice better than NOEMA data for MCG+08$-$11$-$11 and 50 times better than IRAM data for NGC~7469) but also for isolating the flux coming from the compact structure in the core thanks to the $<0$\farcs1 resolution. 

Interestingly, both simultaneous monitoring campaigns presented in \citep{behar20} and \citep{petrucci23}, as well as our results for IC~4329A, while having different measured mm/X-ray ratios, showed a similar scatter in the variations of this ratio ($\sim$0.12~dex). In all cases, this scatter is a factor $\sim$2 lower than that found by \citet{ricci23} for the non-simultaneous mm and X-ray data. Thus, we can argue that half of the scatter in \citet{ricci23} is due to the non-simultaneity of the mm and X-ray observations, while the rest of the scatter could be intrinsic, and associated with the different physical drivers of variability on the short (daily) timescale, or due to the possible time lag between X-rays and mm. However, unlike monitoring IC~4329A, a smaller scatter in the NGC\,7469 and MCG data+08$-$11$-$11 observations may also be related to the contribution from extended non-variable mm radiation, which is why a direct comparison of the scatters is not entirely correct. This once again shows the importance of using high spatial resolution mm observations.

While no clear evidence of correlated mm/X-ray variability was found for NGC\,7469 and MCG+08$-$11$-$11, both AGN showed significant variations in both bands during the monitoring campaigns: $F_{\rm var}^{\rm X-ray} \approx 25$\% and $F_{\rm var}^{\rm mm} \approx 13$\% for NGC~7469 (with the maximum amplitude of a factor of 2 between days), and $F_{\rm var}^{\rm X-ray} \approx 7$\% and $F_{\rm var}^{\rm mm} \approx 2$\%\footnote{Note here that fractional uncertainty of the NOEMA observation is comparable, $\sim$2\%.} for MCG+08$-$11$-$11. In both cases, the X-ray fractional variability was larger than the one in the mm band. However, our data for IC~4329A showed the opposite: $F_{\rm var}^{\rm mm}$ is approximately twice as large as $F_{\rm var}^{\rm X-ray}$. Also, the mm flux of IC~4329A showed a significant flare, during which the flux increased by a factor $\sim$3 within 4--5 days, which was not observed before in other RQ AGN. Due to the limited amount of data, it is still challenging to define whether the observed variability of IC~4329A is a specific characteristic of the source or is typical but not observed in the existing mm observations of NGC~7469 and MCG+08$-$11$-$11 for some reasons. In any case, it is evident that further high spatial resolution variability studies in the mm band are necessary to identify, at the very least, the characteristics of mm variability in RQ AGN.

\subsection{Extended mm structure}

As discussed in Sect.~\ref{mm} and shown in Fig.~\ref{fig:map}, high spatial resolution ALMA observations of IC~4329A revealed a distant structure besides the bright, variable unresolved core in the observed mm region. Due to the faintness of the extended mm structure (signal-to-noise ratio is $\sim$ 20), we cannot examine its properties in detail, but our observations show that its flux is constant at $\sim$0.6~mJy within the errors during the 10-day campaign. Since the bright variable core and the faint extended structure are clearly resolved in the ALMA maps, the linear distance between them projected onto the celestial plane is of the order of 10~pc. 

A non-AGN origin of the extended structure is not plausible. Given the edge-on orientation of the host galaxy of IC~4329A, the observed emission may originate in structures within the galaxy outside the AGN, like HII regions or supernova remnants. Nonetheless, the mm observations from the literature suggest that none of them can yield substantial flux. The anticipated flux from HII regions at redshift $z=0.016$ is merely a few $\mu$Jy when scaling data from \citep[][]{HII}. Similarly, the expected flux from recent supernovae is $<$ $\mu$Jy, as derived from the SN~1987A \citep[][]{sn1987a} and Cas A \citep{casa2,casa1} data. 

The extended structure is possibly associated with the radio jet of IC~4329A, which is detected at lower frequencies \citep[see][]{Inoue18}. The VLA images with a resolution of 1\farcs5 $\times$ 0\farcs9 at 1.5~GHz and 1\farcs3 $\times$ 1\farcs2 at 4.9~GHz from \cite{Unger1987} revealed a bright radio flux. However, the jet extends $\sim$ 6$''$ to the west -- $180^\circ$ opposite to our current mm band observations. Moreover, according to \citet{Inoue18}, the expected mm emission from the jet is considerably higher. Extrapolating from the radio data yields a flux density of $\sim$3.3~mJy at 100 GHz, which is five times higher than the observed emission.

It turned out that the observed emission in the extended mm structure closely matches the expected emission from the photoionized gas in the ionisation cone of the narrow-line region. In this region, mm emission is anticipated as a result of free--free emission \citep[see][]{panessa19, baskin21}. Previous studies on IC~4329A have identified features of the ionisation cone, which align with the extended low-frequency radio jet observed in VLA data \citep{colbert96,Thomas2017}. Following \citet{baskin21}, we estimate the mm flux using the relation $\log \nu L_{\rm 100GHz} = \log L_{\rm [O III]} - 3.36$, where the [OIII] flux $F_{\rm[O III]} = 2.34\cdot10^{-13}$ and the factor 9.5 intrinsic reddening factor are from \citet{bentz4329}. We find the apparent flux $F_{\rm 100GHz} \approx 1$~mJy. To provide a conclusive assessment, checking the spectral slope is essential. Unfortunately, the structure appeared too faint to evaluate its spectral parameters, emphasizing the need for deeper mm observations.

\subsection{mm variability and its origin} \label{mmvar}

One of the key findings of the monitoring is the remarkable variability exhibited by the compact mm source in IC~4329A. Previous studies lacked mm monitoring data for IC~4329A, with only limited ALMA observations. \citet{imanashi16} reported a continuum flux of 13.0~mJy at 260 GHz with a $1\times0$\farcs5 beam in April 2014. Subsequent observations in October 2016 published by \citet{Inoue18} across the 90.5--231~GHz range revealed a frequency-dependent flux, ranging from 8 to $\sim$5~mJy, with spatial resolution between 0.45 and 0.14$''$. Thus, during 10 days of our ALMA monitoring, we captured the full historical range of flux variations, with a flux change of 2.6 within just 4--5 days. 

The observed mm changes are quite surprising. Traditionally, the largest amplitude of variability in non-blazar AGN on timescales less than a month has been observed in the X-ray range, as demonstrated by numerous multiwavelength variability campaigns \citep[e.g.][]{edelson00,edelson15}. However, during our campaign, IC~4329A exhibited more variability in the mm band than in X-rays, as discussed in our comparison of $F_{\rm var}$ in Sect.~\ref{lit}. Based on the factor $\sim$3 variability, we can rule out two of the most stable mechanisms of the mm origin -- thermal dust and free--free emission. 

The dust contribution in IC~4329A is constrained to be at least two orders of magnitude lower than the observed flux at 100~GHz (Fig.~\ref{fig:evolution}), less than 1.6 $\times \ 10^{-2}$~mJy \citep{Inoue18}. What is more important, the observed mm spectral slope $\sim -0.2$ is inconsistent with the dust scenario, where a spectral index of $\sim$3.5 would be expected \citep[e.g.][]{Mullaney11}. The estimated brightness temperature $T_{\rm b}$ from the mm flux density following Eq.~(1.34) from \citet{wilson_book}: $T_{\rm b} \approx 200 \pm 100$~K is also higher than predicted from the SED fitting by \cite{Mehdipour18}. 

In some AGN such as NGC\,1068, a significant fraction of mm emission is described by the free--free emission coming from the winds or X-ray-heated disk gas \citep{Gallimore04, inoue20}. While the observed spectral slope in IC~4329A is consistent with that expected for free--free processes, the flux density and its rapid variations exclude this mechanism. \citet{laor08} demonstrated that the contribution of thermal free--free emission from hot X-ray plasma in the corona is negligible. The free--free emission can originate from the dusty torus in AGN. The torus size measured in IC~4329A is of the order of a few hundred light days \citep[ld; 0.15--0.19~pc][]{kishimoto11,grav23}, and as \citet{baskin21} suggested, the expected variations of the free--free emission originating there should occur on a yearly timescale rather than within days.

Therefore, the preferable mm emission mechanism is non-thermal synchrotron produced in a compact region: since the flux changed by more than 3$\sigma$ from one day to another, we can assume the mm source size is smaller than 1~ld ($\sim$0.0008pc). The observed nearly flat spectrum with a slope of $-0.23$ suggests that the observed frequency is near the turnover between optically thick ($\alpha = 2.5$) and optically thin ($\alpha < -0.5$) regime. Further, let us discuss the estimation of the synchrotron source size and the emission origin.

\subsection{Synchrotron emission}

\begin{figure}
    \centering
    \includegraphics[scale=0.6]{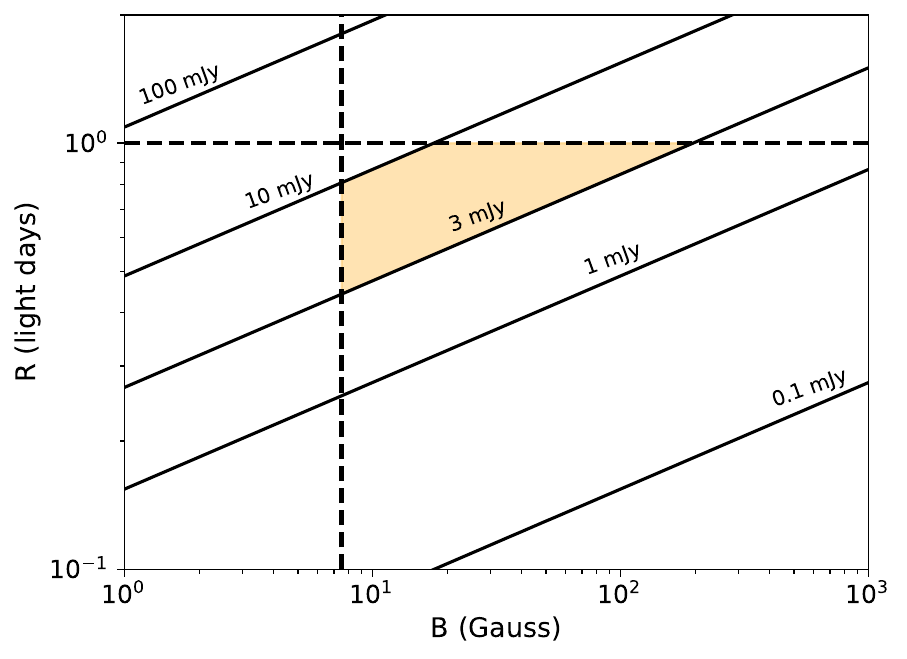}
    \caption{The dependence of the mm flux density at 100~GHz on the source size and magnetic field strength for $z=0.016$. A dashed horizontal line illustrates the region of source sizes $\geq$1~ld. A dashed vertical line delineates the range where the magnetic field $\leq B_{\text{lim}}$. The region between the refined values of $R$ and $B$ for the optically thick synchrotron regime, and the observed flux of IC~4329A is highlighted in yellow.}
    \label{fig:rb}
\end{figure}

In Sect.~\ref{mmvar}, we estimated the mm emitter size as $\sim$1~ld. This estimation aligns perfectly with the optically thick synchrotron source size derived under the assumption of equipartition between magnetic energy density and photon energy density, as introduced by \citet{laor08}:
\begin{equation} 
R = 5.6\times10^2 L_{\rm 100GHz}^{0.4} L_{\rm bol}^{0.1}  \nu^{-1}  \ \ {\rm ld}, 
\end{equation}
where $L_{\rm 100GHz}$ is the mm luminosity in units of $10^{30}$ erg s$^{-1}$ Hz$^{-1}$, $L_{\rm bol}$ is bolometric luminosity in units of $10^{30}$ erg s$^{-1}$, and $\nu$ is the observed frequency in GHz. 
Using the bolometric luminosity calculated in Sect.~\ref{evo} we obtained the source size $1.0\pm0.1$~ld. In the more common case, the physical size of a self-absorbed synchrotron source $R$ is the function of the radio flux density and the magnetic field strength $B$ \citep{laor08,petrucci23}:
\begin{equation} 
R \simeq 3.05 \times 10^3 F_{\rm mm}^{0.5}  \nu^{-1.25} B^{0.25} z  \ \ {\rm ld}, 
\end{equation}
where $F_{\rm mm}$ is in mJy, and $z$ is the source redshift. 
In Fig. \ref{fig:rb}, we present this relation as the $R$--$B$ diagram. The mm flux of IC~4329A varied within the range of $\sim$3--10~mJy. Assuming the emission region size does not exceed 1~ld, we estimate the constraint on the permissible maximum magnetic field strength of 200~G for the minimum flux. The minimum magnetic field value can be estimated following \citet{laor08}, assuming that synchrotron cooling dominates other cooling processes. Thus,
\begin{equation}
B > 6.8 \cdot 10^4 t^{-2/3}_{\text{var}} \nu^{-1/3}   \ \ {\rm G},
\end{equation}
where $t_{\text{var}}$ day is the variability timescale in seconds. We obtained the limiting value $B_{\rm lim} = 7.5$~G, leading to $R > 0.4$~ld for the minimum flux. These constraints significantly limit the permissible parameter space of the mm emitter. In particular, if the source flux is 10~mJy, its size must lie within 0.8--1~ld, and the magnetic field must range between 7.5--18~G. We caution that there is additional uncertainty to these numbers as the source is not deep in the optically thick regime at 100~GHz (as the observed spectral slope is $-0.23$), but this should still provide a valid zero-order estimate of the emitter size and magnetic field.

\begin{figure*}
    \centering

    \subfigure[MJD=59461.41]{\includegraphics[width=0.48\linewidth]{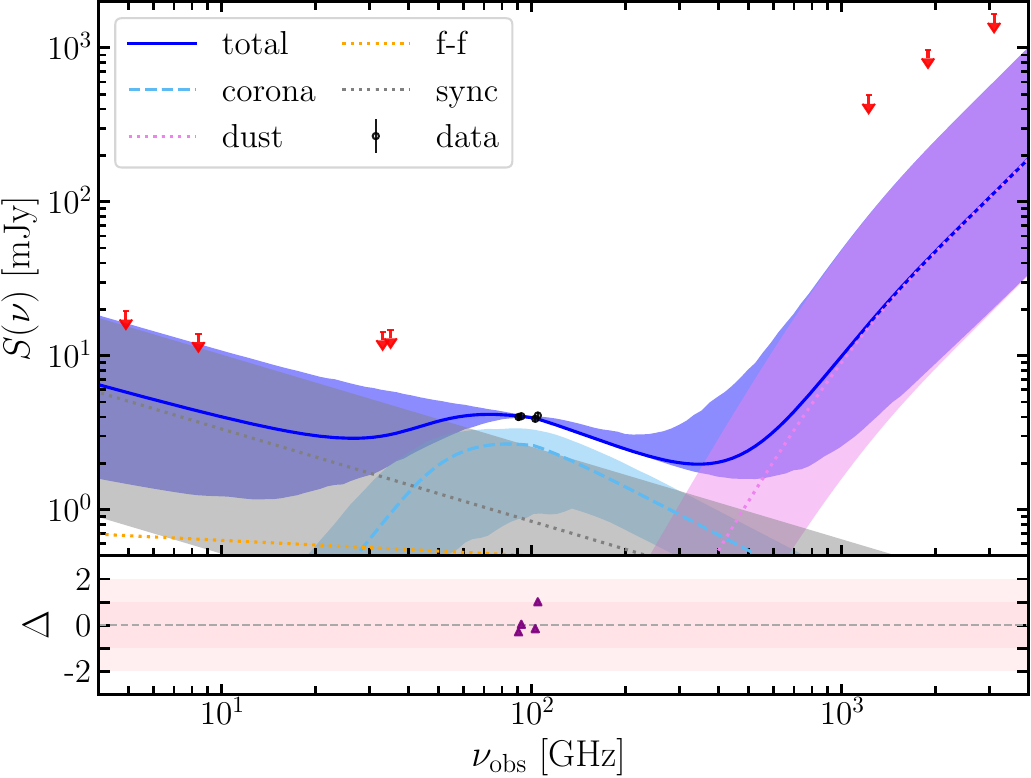}}
   \hfill
    \subfigure[MJD=59464.32]{\includegraphics[width=0.48\linewidth]{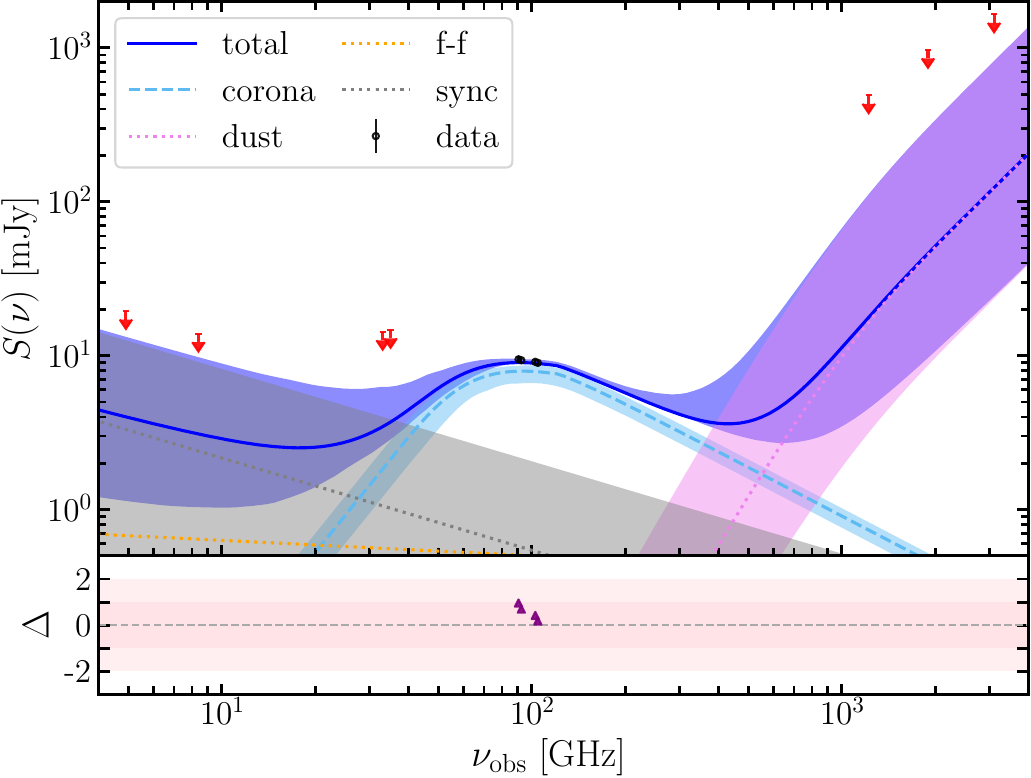}}


    \caption{Multifrequency SED fitting of IC~4329A during the flux minimum and maximum. 
    The black points represent the high-resolution ALMA data (this work), whereas the red arrows are the fluxes from low-resolution radio and infrared observations \citep[][and references therein]{Inoue18}, which in the context of the fitting procedure represent strict upper limits to the total flux coming from the nuclear region probed by ALMA.
    The different emission components in the model and their 1$\sigma$ confidence intervals are marked, together with the total emission. The coronal component dominates the 100~GHz emission, and the diffuse synchrotron and dust emission components are poorly constrained. The bottom subpanels show the fitting residuals. }
    \label{fig:fit}
    
\end{figure*}


Further, let us discuss several possible synchrotron mechanisms of mm emission origin following \citet{Kawamuro22}. There we address a more detailed modelling that does not require the assumption of an optically thick spectrum.


\subsubsection{Synchrotron emission from outflow-driven shocks}
One of the possible explanations for the tight mm/X-ray correlation observed in RQ AGN is synchrotron emission from electrons accelerated in the shock produced by an AGN outflow colliding with the surrounding interstellar medium. \citet{Kawamuro22} showed that the expected conversion of the kinetic energy of the outflow into synchrotron emission of the relativistic particles \citep{nims15} is consistent with the results found, specifically in the case of AGN with ultrafast X-ray outflows. 

The kinetic energy of the outflow is derived from the bolometric luminosity of the AGN, and it is anticipated that more luminous AGN would expel more energetic outflows \citep[see][for references]{Kawamuro22}. Thus, a correlation between the mm/X-ray ratio and the Eddington ratio is expected. \citet{Kawamuro22} reported the absence of such a correlation for the AGN sample. In IC~4329A, where the ultra-fast outflow was found \citep{tombesi12,tortosa23}, we also did not find any relation between the mm/X-ray ratio and $\lambda_{\rm Edd}$ for either $\tau = 0$ or $-11.5$ days. Nevertheless, in the case of $\tau = +15.5$ days, a faint correlation can be assumed with the Pearson coefficient of $\sim$0.65 (see Fig.~\ref{fig:evolution}). However, $+15.5$~days time lag seems inconsistent, as \citet{tombesi12} and \citet{tortosa23} discovered the outflow location in the range of 1.5--12~ld from the core. Moreover, 1.5--12~ld scales of the outflow exceed the expected size of the mm source and, as we show further, should produce optically thin emission at 100~GHz (see details in Sect.~\ref{corona}). 


\subsubsection{Synchrotron emission from jet} \label{jet}

\citet{Kawamuro22} investigated the potential contribution of the jet in the mm emission and found no discernible difference between type~1 and type~2 AGN, suggesting that the jet does not dominate the mm band. However, in IC~4329A, the jet still appears to contribute. As previously noted, extrapolating data from \citet{Inoue18} yielded a mm flux from the jet at 100~GHz of $\sim$3.3~mJy. Interestingly, the source flux has never been observed below this value, implying a significant contribution from the jet in the flux minimum. 

However, the jet is unlikely to contribute to the mm variability significantly. 
The main argument here is that the observed radio jet in IC~4329A exhibits a spectral slope of approximately $-0.59 \pm 0.09$ \citep[in the 1.4--43.3~GHz range][]{Inoue18}. The spectral slope of the mm emission is much flatter and never exhibits a spectral index as negative as the jet component. Furthermore, there is no indication that the spectral slope becomes more negative as the flux increases (see Fig.~\ref{fig:ALMA_freq}). Additionally, the daily-scale variability seems inconsistent with the activity of the radio jet with the typical scale exceeding several kpc.  However, it is worth noting that rapid variability may originate from a compact jet even in non-blazar AGN \citep[e.g.][]{pictor_a}, thus this requires additional studies.
These findings suggest that the contribution of the jet to the mm emission is limited and possibly exhibits slower changes compared to the observed mm variability.

At present, the jet scenario is not favoured. However, determining the exact contribution of the jet to the mm flux and whether this contribution is variable remains an intriguing open question, necessitating simultaneous observations at lower radio frequencies and in the mm band.

\subsubsection{Synchrotron emission from X-ray corona} \label{corona}

As shown by \citet{behar18}, the mm excess observed in many RQ AGN cannot be extrapolated from the radio data. In Sect.~\ref{jet}, we demonstrated that, in the case of IC~4329A, the minimum of the flux aligns well with the radio observations. However, this cannot fully account for the observed variability and the mm excess in IC~4329A during its maximum. 

\citet{Inoue14} proposed a scenario in which non-thermal relativistic electrons produce mm emission in the magnetised X-ray corona. The model was further investigated by \citet{Inoue18}, predicting that synchrotron emission peaks in the mm band, at $\sim$100--300~GHz, due to synchrotron self-absorption (SSA). 
The synchrotron self-absorption scenario was also preferred by \citet[][see their Fig. 25]{Kawamuro22}. For IC~4329A, we found a flatter spectral slope than the average of the sample in \citep{Kawamuro22} ($\alpha_{\rm Kawamuro}^{\rm ave} = -0.5\pm1.2$), which could be explained by observations conducted at 100~GHz, closer to the SSA peak, and not at 230~GHz. 

In this scenario, the synchrotron emissivity is a function of the electron spectrum and the magnetic field strength and size of the corona. Thus, to produce the observed rapid variability on timescales of days, fast changes in the physical properties of the corona are required. To explore this more carefully, we introduce a corona emission model based on \cite{Inoue18} (see details in del Palacio et al. 2024, in prep.). This model has many parameters to characterise the coronal properties and the non-thermal electron distribution in it. Specifically, the parameters include the temperature, electron density, and size of the corona, the fraction of energy in the magnetic field and non-thermal electrons, and the spectral index of the relativistic electron energy distribution. Most of these parameters affect the SED in a similar way, increasing both the peak frequency and the peak flux (due to increasing SSA opacity and synchrotron emissivity). The only parameter that increases the flux while decreasing the peak frequency is the size of the corona. As usually done, we parameterize this as $R_\mathrm{c} = r_\mathrm{c} R_\mathrm{Sch}$, where $R_\mathrm{Sch} \propto M_\mathrm{BH}$ is the Schwarzschild radius. Thus, if the flux varies and the spectral index remains flat, the size must change. Otherwise, the emission would get a more positive spectral index when it is brighter, which is not observed (Fig. \ref{fig:ALMA_freq}). 

Not all parameters allow the reproduction of the observed flux density and flat spectral index. For example, fixing $r_\mathrm{c}=40$ for the brightest epoch and varying the remaining parameters, the model fails to reproduce the spectral index and requires unphysical parameters (100 times more energy in non-thermal electrons than in thermal ones). 
This is because the only factor that can reduce the high SSA opacity without leading to a low flux density is the corona size, with a larger size yielding a more diluted electron population in the corona. 
For this reason, we allow the parameters $r_\mathrm{c}$ and $\delta = U_\mathrm{nt,e}/U_\mathrm{th,e}$ (fraction of the energy in non-thermal electrons w.r.t. thermal electrons) to vary, while we fix the remaining parameters of the corona, $kT = 200$ keV, $\tau_\mathrm{T}=0.43$, $\epsilon_B = U_\mathrm{B}/U_\mathrm{th,e} = 10^{-2}$ (fraction of energy in the magnetic field), and $p=2.1$ (relativistic electron distribution); we also fix the dust parameters $\beta = 1.6$ and $\nu_{\tau=1}=800$ GHz, and the spectral index of the diffuse synchrotron emission to $\alpha=-0.6$. 
We note that the diffuse components are very poorly constrained due to the lack of observations with comparable resolution at other frequencies. Given that low-resolution observations probe larger volumes in the galaxy and thus capture more emission, the fluxes from archival low-resolution observations \citep[][and references therein]{Inoue18} represent strict upper limits to the flux coming from the inner region probed by the high-resolution ALMA observations.
In Fig.~\ref{fig:fit}, we present the SED fitting of IC~4329A during two days that show the highest variability. We obtain that the corona expanded from $r_c \approx 116 \pm 54$ ($R_c \approx 0.91$~ld) during the minimum (MJD 59461.4) to $r_c \approx 166 \pm 25$ ($R_c\approx1.3$~ld) during the maximum (MJD 59464.3) with the magnetic field strength varying from 6.6 to 5.6~G and the fraction of non-thermal electrons increasing from $\log \delta = -3.18 \pm 0.3$ to $\log \delta = -2.87 \pm 0.1$. 
The change in the size of the corona of 0.4~ld in 3 days would require a very fast expansion velocity of $\sim 0.1\,c$, which could be consistent with the outflowing corona scenario \citep[e.g.][]{Kylafis2023}.

In summary, our preferred scenario attributes the mm variability and tight X-ray/mm ratio to a compact X-ray corona. However, in this scenario, as well as in the jet and outflow scenarios, no time lag is expected. The 11.5 and 15.5 days time lags pose challenges, as we struggle to envision how the source could accumulate and later reradiate energy in mm or X-rays within such time frames. Treating these time lags as light travel times is not a satisfactory explanation, as it would require the sources to be separated by a distance approximately equivalent to the broad-line region (BLR), which is about 16 ld \citep{bentz4329}. Within our current understanding of the mm source, a zero time lag ($\tau=0$) seems to be more plausible.

Additionally, none of the discussed mechanisms accounts for the 5-day timescale variations observed in X-rays and their absence in mm. The distinct variability patterns over several days hint at the potential for common X-ray and mm variability on longer timescales (weeks), while the daily timescale variations suggest differences in short-term processes between X-rays and mm, possibly involving distinct electron populations.
Furthermore, investigating X-ray variability showing local periodic changes independently holds interest, given that X-rays are generally known for stochastic variability. However, comparing X-ray periods with known timescales in IC~4329A fails to provide insights into their physical origin.

\section{Conclusions}

We observed the RQ unobscured type 1 AGN IC~4329A to investigate the origin and properties of the compact mm emission in RQ AGN. The source was observed daily for 40 days in the 2--10\,keV X-ray band using \xmm, \textit{Swift}, and \textit{NICER}. Within this campaign, we contemporaneously obtained 10 epochs of high-resolution mm observations with daily cadence using ALMA at $\sim$100~GHz frequency. 
Unlike previous contemporaneous mm/X-ray campaigns \citep{behar20,petrucci23}, the outstanding capabilities of ALMA enable us to isolate the compact mm emission from the potential contributions of other components of the active galaxy and to observe the surprising behaviour of the mm emission in the RQ AGN.

The recent results of the tight mm/X-ray correlation in RQ AGN by \citet{ricci23} and \citet{Kawamuro22} suggested the common origin of the mm and X-ray emission most likely in the X-ray corona. Thus, we expected to discover a correlated behaviour of both fluxes on a daily timescale. However, we did not find statistically robust evidence of the correlated variability, and distinct variability patterns were observed in each band.
\begin{itemize}
    \item The observed ratio between X-ray (2--10~keV) and mm (100~GHz) aligns with the tight correlation reported in \citet{Kawamuro22, ricci23}. Moreover, our simultaneous observations indicate a reduced scatter in this ratio, measuring $\sim$0.1~dex instead of the 0.22~dex \citep{ricci23}. 
    This suggests that at least half of the previously observed scatter is due to the non-simultaneity of the observations, while the rest may arise from intrinsic processes in the AGN.  
    \item During the 10-day campaign, the compact mm emission exhibited significant variability by a factor of three within 4--5 days, with an amplitude surprisingly larger than in X-rays. This rules out the heated dust and thermal free--free emission origin, indicating a synchrotron origin for the mm radiation. The in-band spectral slope $-0.23 \pm 0.18$ suggests that the observed emission is close to the SSA turnover frequency. The SED fitting modelling suggests a source size of $\sim$1~ld ($\sim$120 gravitational radii), consistent with size estimations from the variability timescale, and a magnetic field strength of $\sim$5--7~G. The observed variability and spectral indices require the emitter size to change in time. 
    \item The combined analysis of the 40-day X-ray (2--10~keV) and 10-day mm (100~GHz) light curves suggests no reliable time lag. 
    Additionally, the tentative oscillations on the 5-day timescale observed in X-rays at the 95\% confidence level are not evident in the mm variability. 
    \item The high resolution of ALMA (up to $\sim$0\farcs05) allowed us to resolve the variable bright core together with the faint and non-variable mm structure 10~pc away from the nucleus. The measured flux and the position of the non-variable component indicate that its emission likely originates in the ionization cone of the narrow-line region.

\end{itemize}

The evidence (flux density, flat/negative spectral index, daily variability) strongly suggests that the observed mm radiation has a non-thermal synchrotron nature and originates from a compact source. This is best consistent with the idea that, as suggested by \citet{Inoue14} and mm/X-ray correlations, the mm emission is produced in a compact X-ray corona. 
Future simultaneous radio/mm/X-ray observations, possibly covering more ALMA bands and longer timescales, as well as mm polarization observations, will help to elucidate the physical processes in the compact mm emitting region observed in RQ AGNs.


\begin{acknowledgements}
We are grateful to the ALMA, NICER and XMM-Newton teams for their invaluable help in coordinating the joint mm/ALMA observations. We sincerely thank Ari Laor for the comments and fruitful discussion.
ES acknowledges ANID BASAL project FB210003 and Gemini ANID ASTRO21-0003. 
CR acknowledges support from Fondecyt Regular grant 1230345 and ANID BASAL project FB210003.
SdP and SA gratefully acknowledge funding from the European Research Council (ERC) under the European Union's Horizon 2020 research and innovation programme (grant agreement No 789410, PI: S. Aalto).
We gratefully acknowledge funding for this work by ANID through the Millennium Science Initiative Program - ICN12\_009 (FEB), CATA-BASAL - FB210003 (FEB), and FONDECYT Regular - 1200495 (FEB).
LCH was supported by the National Science Foundation of China (11991052, 12011540375, 12233001), the National Key R\&D Program of China (2022YFF0503401), and the China Manned Space Project (CMS-CSST-2021-A04, CMS-CSST-2021-A06).
The material is based upon work supported by NASA under award number 80GSFC21M0002 (ML).
FT acknowledges funding from the European Union - Next Generation EU, PRIN/MUR 2022 (2022K9N5B4). 
The National Radio Astronomy Observatory is a facility of the National
Science Foundation operated under cooperative agreement by Associated
Universities, Inc.

\end{acknowledgements}

%
  \bibliographystyle{aa} 
  \bibliography{biblio} 

\begin{thebibliography}{73}
\expandafter\ifx\csname natexlab\endcsname\relax\def\natexlab#1{#1}\fi

\bibitem[{{Alhosani} {et~al.}(2022){Alhosani}, {Gelfand}, {Zaw}, {Laor}, {Behar}, {Chen}, \& {Wrzosek}}]{alhosani22}
{Alhosani}, A., {Gelfand}, J.~D., {Zaw}, I., {et~al.} 2022, \apj, 936, 73

\bibitem[{{Arzoumanian} {et~al.}(2014){Arzoumanian}, {Gendreau}, {Baker}, {Cazeau}, {Hestnes}, {Kellogg}, {Kenyon}, {Kozon}, {Liu}, {Manthripragada}, {Markwardt}, {Mitchell}, {Mitchell}, {Monroe}, {Okajima}, {Pollard}, {Powers}, {Savadkin}, {Winternitz}, {Chen}, {Wright}, {Foster}, {Prigozhin}, {Remillard}, \& {Doty}}]{Arzoumanian14}
{Arzoumanian}, Z., {Gendreau}, K.~C., {Baker}, C.~L., {et~al.} 2014, in Society of Photo-Optical Instrumentation Engineers (SPIE) Conference Series, Vol. 9144, Space Telescopes and Instrumentation 2014: Ultraviolet to Gamma Ray, ed. T.~{Takahashi}, J.-W.~A. {den Herder}, \& M.~{Bautz}, 914420

\bibitem[{{Baldi} {et~al.}(2015){Baldi}, {Behar}, {Laor}, \& {Horesh}}]{Baldi15}
{Baldi}, R.~D., {Behar}, E., {Laor}, A., \& {Horesh}, A. 2015, \mnras, 454, 4277

\bibitem[{{Baskin} \& {Laor}(2021)}]{baskin21}
{Baskin}, A. \& {Laor}, A. 2021, \mnras, 508, 680

\bibitem[{{Behar} {et~al.}(2015){Behar}, {Baldi}, {Laor}, {Horesh}, {Stevens}, \& {Tzioumis}}]{behar15}
{Behar}, E., {Baldi}, R.~D., {Laor}, A., {et~al.} 2015, \mnras, 451, 517

\bibitem[{{Behar} {et~al.}(2020){Behar}, {Kaspi}, {Paubert}, {Billot}, {Peretz}, {Baldi}, {Laor}, {Kaastra}, \& {Mehdipour}}]{behar20}
{Behar}, E., {Kaspi}, S., {Paubert}, G., {et~al.} 2020, \mnras, 491, 3523

\bibitem[{{Behar} {et~al.}(2018){Behar}, {Vogel}, {Baldi}, {Smith}, \& {Mushotzky}}]{behar18}
{Behar}, E., {Vogel}, S., {Baldi}, R.~D., {Smith}, K.~L., \& {Mushotzky}, R.~F. 2018, \mnras, 478, 399

\bibitem[{{Bentz} {et~al.}(2023){Bentz}, {Onken}, {Street}, \& {Valluri}}]{bentz4329}
{Bentz}, M.~C., {Onken}, C.~A., {Street}, R., \& {Valluri}, M. 2023, \apj, 944, 29

\bibitem[{{Burrows} {et~al.}(2005){Burrows}, {Hill}, {Nousek}, {Kennea}, {Wells}, {Osborne}, {Abbey}, {Beardmore}, {Mukerjee}, {Short}, {Chincarini}, {Campana}, {Citterio}, {Moretti}, {Pagani}, {Tagliaferri}, {Giommi}, {Capalbi}, {Tamburelli}, {Angelini}, {Cusumano}, {Br{\"a}uninger}, {Burkert}, \& {Hartner}}]{Burrows05}
{Burrows}, D.~N., {Hill}, J.~E., {Nousek}, J.~A., {et~al.} 2005, \ssr, 120, 165

\bibitem[{{Chen} {et~al.}(2023){Chen}, {Kharb}, {Sasikumar}, {Nandi}, {Berton}, {Jarvela}, {Laor}, {Behar}, {Foschini}, {Vietri}, {Gu}, {La Mura}, {Crepaldi}, \& {Zhou}}]{Chen23}
{Chen}, S., {Kharb}, P., {Sasikumar}, S., {et~al.} 2023, arXiv e-prints, arXiv:2312.13351

\bibitem[{{Colbert} {et~al.}(1996){Colbert}, {Baum}, {Gallimore}, {O'Dea}, {Lehnert}, {Tsvetanov}, {Mulchaey}, \& {Caganoff}}]{colbert96}
{Colbert}, E. J.~M., {Baum}, S.~A., {Gallimore}, J.~F., {et~al.} 1996, \apjs, 105, 75

\bibitem[{{Donnan} {et~al.}(2021){Donnan}, {Horne}, \& {Hern{\'a}ndez Santisteban}}]{pyroa}
{Donnan}, F.~R., {Horne}, K., \& {Hern{\'a}ndez Santisteban}, J.~V. 2021, \mnras, 508, 5449

\bibitem[{{Edelson} {et~al.}(2015){Edelson}, {Gelbord}, {Horne}, {McHardy}, {Peterson}, {Ar{\'e}valo}, {Breeveld}, {De Rosa}, {Evans}, {Goad}, {Kriss}, {Brandt}, {Gehrels}, {Grupe}, {Kennea}, {Kochanek}, {Nousek}, {Papadakis}, {Siegel}, {Starkey}, {Uttley}, {Vaughan}, {Young}, {Barth}, {Bentz}, {Brewer}, {Crenshaw}, {Dalla Bont{\`a}}, {De Lorenzo-C{\'a}ceres}, {Denney}, {Dietrich}, {Ely}, {Fausnaugh}, {Grier}, {Hall}, {Kaastra}, {Kelly}, {Korista}, {Lira}, {Mathur}, {Netzer}, {Pancoast}, {Pei}, {Pogge}, {Schimoia}, {Treu}, {Vestergaard}, {Villforth}, {Yan}, \& {Zu}}]{edelson15}
{Edelson}, R., {Gelbord}, J.~M., {Horne}, K., {et~al.} 2015, \apj, 806, 129

\bibitem[{{Edelson} {et~al.}(2000){Edelson}, {Koratkar}, {Nandra}, {Goad}, {Peterson}, {Collier}, {Krolik}, {Malkan}, {Maoz}, {O'Brien}, {Shull}, {Vaughan}, \& {Warwick}}]{edelson00}
{Edelson}, R., {Koratkar}, A., {Nandra}, K., {et~al.} 2000, \apj, 534, 180

\bibitem[{{Evans} {et~al.}(2009){Evans}, {Beardmore}, {Page}, {Osborne}, {O'Brien}, {Willingale}, {Starling}, {Burrows}, {Godet}, {Vetere}, {Racusin}, {Goad}, {Wiersema}, {Angelini}, {Capalbi}, {Chincarini}, {Gehrels}, {Kennea}, {Margutti}, {Morris}, {Mountford}, {Pagani}, {Perri}, {Romano}, \& {Tanvir}}]{Evans09}
{Evans}, P.~A., {Beardmore}, A.~P., {Page}, K.~L., {et~al.} 2009, \mnras, 397, 1177

\bibitem[{{Gabriel} {et~al.}(2004){Gabriel}, {Denby}, {Fyfe}, {Hoar}, {Ibarra}, {Ojero}, {Osborne}, {Saxton}, {Lammers}, \& {Vacanti}}]{Gabriel2004}
{Gabriel}, C., {Denby}, M., {Fyfe}, D.~J., {et~al.} 2004, in Astronomical Society of the Pacific Conference Series, Vol. 314, Astronomical Data Analysis Software and Systems (ADASS) XIII, ed. F.~{Ochsenbein}, M.~G. {Allen}, \& D.~{Egret}, 759

\bibitem[{{Gallimore} {et~al.}(2004){Gallimore}, {Baum}, \& {O'Dea}}]{Gallimore04}
{Gallimore}, J.~F., {Baum}, S.~A., \& {O'Dea}, C.~P. 2004, \apj, 613, 794

\bibitem[{{Gaskell} \& {Sparke}(1986)}]{gaskell86}
{Gaskell}, C.~M. \& {Sparke}, L.~S. 1986, \apj, 305, 175

\bibitem[{{Gehrels} {et~al.}(2004){Gehrels}, {Chincarini}, {Giommi}, {Mason}, {Nousek}, {Wells}, {White}, {Barthelmy}, {Burrows}, {Cominsky}, {Hurley}, {Marshall}, {M{\'e}sz{\'a}ros}, {Roming}, {Angelini}, {Barbier}, {Belloni}, {Campana}, {Caraveo}, {Chester}, {Citterio}, {Cline}, {Cropper}, {Cummings}, {Dean}, {Feigelson}, {Fenimore}, {Frail}, {Fruchter}, {Garmire}, {Gendreau}, {Ghisellini}, {Greiner}, {Hill}, {Hunsberger}, {Krimm}, {Kulkarni}, {Kumar}, {Lebrun}, {Lloyd-Ronning}, {Markwardt}, {Mattson}, {Mushotzky}, {Norris}, {Osborne}, {Paczynski}, {Palmer}, {Park}, {Parsons}, {Paul}, {Rees}, {Reynolds}, {Rhoads}, {Sasseen}, {Schaefer}, {Short}, {Smale}, {Smith}, {Stella}, {Tagliaferri}, {Takahashi}, {Tashiro}, {Townsley}, {Tueller}, {Turner}, {Vietri}, {Voges}, {Ward}, {Willingale}, {Zerbi}, \& {Zhang}}]{Gehrels04}
{Gehrels}, N., {Chincarini}, G., {Giommi}, P., {et~al.} 2004, \apj, 611, 1005

\bibitem[{{Gendreau} {et~al.}(2016){Gendreau}, {Arzoumanian}, {Adkins}, {Albert}, {Anders}, {Aylward}, {Baker}, {Balsamo}, {Bamford}, {Benegalrao}, {Berry}, {Bhalwani}, {Black}, {Blaurock}, {Bronke}, {Brown}, {Budinoff}, {Cantwell}, {Cazeau}, {Chen}, {Clement}, {Colangelo}, {Coleman}, {Coopersmith}, {Dehaven}, {Doty}, {Egan}, {Enoto}, {Fan}, {Ferro}, {Foster}, {Galassi}, {Gallo}, {Green}, {Grosh}, {Ha}, {Hasouneh}, {Heefner}, {Hestnes}, {Hoge}, {Jacobs}, {J{\o}rgensen}, {Kaiser}, {Kellogg}, {Kenyon}, {Koenecke}, {Kozon}, {LaMarr}, {Lambertson}, {Larson}, {Lentine}, {Lewis}, {Lilly}, {Liu}, {Malonis}, {Manthripragada}, {Markwardt}, {Matonak}, {Mcginnis}, {Miller}, {Mitchell}, {Mitchell}, {Mohammed}, {Monroe}, {Montt de Garcia}, {Mul{\'e}}, {Nagao}, {Ngo}, {Norris}, {Norwood}, {Novotka}, {Okajima}, {Olsen}, {Onyeachu}, {Orosco}, {Peterson}, {Pevear}, {Pham}, {Pollard}, {Pope}, {Powers}, {Powers}, {Price}, {Prigozhin}, {Ramirez}, {Reid}, {Remillard}, {Rogstad}, {Rosecrans}, {Rowe}, {Sager}, {Sanders},
  {Savadkin}, {Saylor}, {Schaeffer}, {Schweiss}, {Semper}, {Serlemitsos}, {Shackelford}, {Soong}, {Struebel}, {Vezie}, {Villasenor}, {Winternitz}, {Wofford}, {Wright}, {Yang}, \& {Yu}}]{Gendreau16}
{Gendreau}, K.~C., {Arzoumanian}, Z., {Adkins}, P.~W., {et~al.} 2016, in Society of Photo-Optical Instrumentation Engineers (SPIE) Conference Series, Vol. 9905, Space Telescopes and Instrumentation 2016: Ultraviolet to Gamma Ray, ed. J.-W.~A. {den Herder}, T.~{Takahashi}, \& M.~{Bautz}, 99051H

\bibitem[{{Gendreau} {et~al.}(2012){Gendreau}, {Arzoumanian}, \& {Okajima}}]{Gendreau12}
{Gendreau}, K.~C., {Arzoumanian}, Z., \& {Okajima}, T. 2012, in Society of Photo-Optical Instrumentation Engineers (SPIE) Conference Series, Vol. 8443, Space Telescopes and Instrumentation 2012: Ultraviolet to Gamma Ray, ed. T.~{Takahashi}, S.~S. {Murray}, \& J.-W.~A. {den Herder}, 844313

\bibitem[{{Gravity Collaboration} {et~al.}(2023){Gravity Collaboration}, {Amorim}, {Bourdarot}, {Brandner}, {Cao}, {Cl{\'e}net}, {Davies}, {de Zeeuw}, {Dexter}, {Drescher}, {Eckart}, {Eisenhauer}, {Fabricius}, {F{\"o}rster Schreiber}, {Garcia}, {Genzel}, {Gillessen}, {Gratadour}, {H{\"o}nig}, {Kishimoto}, {Lacour}, {Lutz}, {Millour}, {Netzer}, {Ott}, {Paumard}, {Perraut}, {Perrin}, {Peterson}, {Petrucci}, {Pfuhl}, {Prieto}, {Rouan}, {Santos}, {Shangguan}, {Shimizu}, {Sternberg}, {Straubmeier}, {Sturm}, {Tacconi}, {Tristram}, {Widmann}, \& {Woillez}}]{grav23}
{Gravity Collaboration}, {Amorim}, A., {Bourdarot}, G., {et~al.} 2023, \aap, 669, A14

\bibitem[{{Guedel} \& {Benz}(1993)}]{guedel93}
{Guedel}, M. \& {Benz}, A.~O. 1993, \apjl, 405, L63

\bibitem[{{Hunter} {et~al.}(2023){Hunter}, {Indebetouw}, {Brogan}, {Berry}, {Chang}, {Francke}, {Geers}, {G{\'o}mez}, {Hibbard}, {Humphreys}, {Kent}, {Kepley}, {Kunneriath}, {Lipnicky}, {Loomis}, {Mason}, {Masters}, {Maud}, {Muders}, {Sabater}, {Sugimoto}, {Sz{\H{u}}cs}, {Vasiliev}, {Videla}, {Villard}, {Williams}, {Xue}, \& {Yoon}}]{alma_pipeline}
{Hunter}, T.~R., {Indebetouw}, R., {Brogan}, C.~L., {et~al.} 2023, \pasp, 135, 074501

\bibitem[{{Imanishi} {et~al.}(2016){Imanishi}, {Nakanishi}, \& {Izumi}}]{imanashi16}
{Imanishi}, M., {Nakanishi}, K., \& {Izumi}, T. 2016, \aj, 152, 218

\bibitem[{{Inoue} \& {Doi}(2014)}]{Inoue14}
{Inoue}, Y. \& {Doi}, A. 2014, \pasj, 66, L8

\bibitem[{{Inoue} \& {Doi}(2018)}]{Inoue18}
{Inoue}, Y. \& {Doi}, A. 2018, \apj, 869, 114

\bibitem[{{Inoue} {et~al.}(2020){Inoue}, {Khangulyan}, \& {Doi}}]{inoue20}
{Inoue}, Y., {Khangulyan}, D., \& {Doi}, A. 2020, \apjl, 891, L33

\bibitem[{{Jansen} {et~al.}(2001){Jansen}, {Lumb}, {Altieri}, {Clavel}, {Ehle}, {Erd}, {Gabriel}, {Guainazzi}, {Gondoin}, {Much}, {Munoz}, {Santos}, {Schartel}, {Texier}, \& {Vacanti}}]{Jansen2001}
{Jansen}, F., {Lumb}, D., {Altieri}, B., {et~al.} 2001, \aap, 365, L1

\bibitem[{{Kawamuro} {et~al.}(2022){Kawamuro}, {Ricci}, {Imanishi}, {Mushotzky}, {Izumi}, {Ricci}, {Bauer}, {Koss}, {Trakhtenbrot}, {Ichikawa}, {Rojas}, {Smith}, {Shimizu}, {Oh}, {den Brok}, {Baba}, {Balokovi{\'c}}, {Chang}, {Kakkad}, {Pfeifle}, {Privon}, {Temple}, {Ueda}, {Harrison}, {Powell}, {Stern}, {Urry}, \& {Sanders}}]{Kawamuro22}
{Kawamuro}, T., {Ricci}, C., {Imanishi}, M., {et~al.} 2022, \apj, 938, 87

\bibitem[{{Kawamuro} {et~al.}(2023){Kawamuro}, {Ricci}, {Mushotzky}, {Imanishi}, {Bauer}, {Ricci}, {Koss}, {Privon}, {Trakhtenbrot}, {Izumi}, {Ichikawa}, {Rojas}, {Smith}, {Shimizu}, {Oh}, {den Brok}, {Baba}, {Balokovi{\'c}}, {Chang}, {Kakkad}, {Pfeifle}, {Temple}, {Ueda}, {Harrison}, {Powell}, {Stern}, {Urry}, \& {Sanders}}]{Kawamuro23}
{Kawamuro}, T., {Ricci}, C., {Mushotzky}, R.~F., {et~al.} 2023, \apjs, 269, 24

\bibitem[{{Kishimoto} {et~al.}(2011){Kishimoto}, {H{\"o}nig}, {Antonucci}, {Barvainis}, {Kotani}, {Tristram}, {Weigelt}, \& {Levin}}]{kishimoto11}
{Kishimoto}, M., {H{\"o}nig}, S.~F., {Antonucci}, R., {et~al.} 2011, \aap, 527, A121

\bibitem[{{Koss} {et~al.}(2022){Koss}, {Ricci}, {Trakhtenbrot}, {Oh}, {den Brok}, {Mej{\'\i}a-Restrepo}, {Stern}, {Privon}, {Treister}, {Powell}, {Mushotzky}, {Bauer}, {Ananna}, {Balokovi{\'c}}, {B{\"a}r}, {Becker}, {Bessiere}, {Burtscher}, {Caglar}, {Congiu}, {Evans}, {Harrison}, {Heida}, {Ichikawa}, {Kamraj}, {Lamperti}, {Pacucci}, {Ricci}, {Riffel}, {Rojas}, {Schawinski}, {Temple}, {Urry}, {Veilleux}, \& {Williams}}]{koss_catalogue}
{Koss}, M.~J., {Ricci}, C., {Trakhtenbrot}, B., {et~al.} 2022, \apjs, 261, 2

\bibitem[{{Kylafis} {et~al.}(2023){Kylafis}, {Reig}, \& {Tsouros}}]{Kylafis2023}
{Kylafis}, N.~D., {Reig}, P., \& {Tsouros}, A. 2023, \aap, 679, A81

\bibitem[{{Laki{\'c}evi{\'c}} {et~al.}(2011){Laki{\'c}evi{\'c}}, {van Loon}, {Patat}, {Staveley-Smith}, \& {Zanardo}}]{sn1987a}
{Laki{\'c}evi{\'c}}, M., {van Loon}, J.~T., {Patat}, F., {Staveley-Smith}, L., \& {Zanardo}, G. 2011, \aap, 532, L8

\bibitem[{{Laor} \& {Behar}(2008)}]{laor08}
{Laor}, A. \& {Behar}, E. 2008, \mnras, 390, 847

\bibitem[{{Li} {et~al.}(2019){Li}, {Shen}, {Brandt}, {Grier}, {Hall}, {Ho}, {Homayouni}, {Horne}, {Schneider}, {Trump}, \& {Starkey}}]{Li19}
{Li}, I-Hsiu, J., {Shen}, Y., {Brandt}, W.~N., {et~al.} 2019, \apj, 884, 119

\bibitem[{{Loinard} {et~al.}(2003){Loinard}, {Lequeux}, {Tilanus}, \& {Lagage}}]{casa1}
{Loinard}, L., {Lequeux}, J., {Tilanus}, R.~P.~T., \& {Lagage}, P.~O. 2003, in Revista Mexicana de Astronomia y Astrofisica Conference Series, Vol.~15, Revista Mexicana de Astronomia y Astrofisica Conference Series, ed. J.~{Arthur} \& W.~J. {Henney}, 267--269

\bibitem[{{Lomb}(1976)}]{lomb}
{Lomb}, N.~R. 1976, \apss, 39, 447

\bibitem[{{Mehdipour} \& {Costantini}(2018)}]{Mehdipour18}
{Mehdipour}, M. \& {Costantini}, E. 2018, \aap, 619, A20

\bibitem[{{Mehdipour} {et~al.}(2017){Mehdipour}, {Kaastra}, {Kriss}, {Arav}, {Behar}, {Bianchi}, {Branduardi-Raymont}, {Cappi}, {Costantini}, {Ebrero}, {Di Gesu}, {Kaspi}, {Mao}, {De Marco}, {Matt}, {Paltani}, {Peretz}, {Peterson}, {Petrucci}, {Pinto}, {Ponti}, {Ursini}, {de Vries}, \& {Walton}}]{Mehdipour17}
{Mehdipour}, M., {Kaastra}, J.~S., {Kriss}, G.~A., {et~al.} 2017, \aap, 607, A28

\bibitem[{{Merloni} \& {Fabian}(2001{\natexlab{a}})}]{merloni01b}
{Merloni}, A. \& {Fabian}, A.~C. 2001{\natexlab{a}}, \mnras, 321, 549

\bibitem[{{Merloni} \& {Fabian}(2001{\natexlab{b}})}]{merloni01a}
{Merloni}, A. \& {Fabian}, A.~C. 2001{\natexlab{b}}, \mnras, 328, 958

\bibitem[{Morlet(1983)}]{Morlet1983}
Morlet, J. 1983, Sampling Theory and Wave Propagation (Springer Berlin Heidelberg), 233–261

\bibitem[{{Mullaney} {et~al.}(2011){Mullaney}, {Alexander}, {Goulding}, \& {Hickox}}]{Mullaney11}
{Mullaney}, J.~R., {Alexander}, D.~M., {Goulding}, A.~D., \& {Hickox}, R.~C. 2011, \mnras, 414, 1082

\bibitem[{{Nims} {et~al.}(2015){Nims}, {Quataert}, \& {Faucher-Gigu{\`e}re}}]{nims15}
{Nims}, J., {Quataert}, E., \& {Faucher-Gigu{\`e}re}, C.-A. 2015, \mnras, 447, 3612

\bibitem[{{Oknyanskii}(1993)}]{okn93}
{Oknyanskii}, V.~L. 1993, Astronomy Letters, 19, 416

\bibitem[{{Oknyansky} \& {Oknyansky}(2022)}]{pymccf}
{Oknyansky}, V. \& {Oknyansky}, R. 2022, {PyMCCF: Python Modernized Cross Correlation Function for reverberation mapping studies}, Astrophysics Source Code Library, record ascl:2212.007

\bibitem[{{Panessa} {et~al.}(2019){Panessa}, {Baldi}, {Laor}, {Padovani}, {Behar}, \& {McHardy}}]{panessa19}
{Panessa}, F., {Baldi}, R.~D., {Laor}, A., {et~al.} 2019, Nature Astronomy, 3, 387

\bibitem[{{Peterson} {et~al.}(2014){Peterson}, {Grier}, {Horne}, {Pogge}, {Bentz}, {De Rosa}, {Denney}, {Martini}, {Sergeev}, {Kaspi}, {Minezaki}, {Zu}, {Kochanek}, {Siverd}, {Shappee}, {Araya Salvo}, {Beatty}, {Bird}, {Bord}, {Borman}, {Che}, {Chen}, {Cohen}, {Dietrich}, {Doroshenko}, {Drake}, {Efimov}, {Free}, {Ginsburg}, {Henderson}, {King}, {Koshida}, {Mogren}, {Molina}, {Mosquera}, {Motohara}, {Nazarov}, {Okhmat}, {Pejcha}, {Rafter}, {Shields}, {Skowron}, {Skowron}, {Valluri}, {van Saders}, \& {Yoshii}}]{Peterson7469}
{Peterson}, B.~M., {Grier}, C.~J., {Horne}, K., {et~al.} 2014, \apj, 795, 149

\bibitem[{{Peterson} {et~al.}(1998){Peterson}, {Wanders}, {Horne}, {Collier}, {Alexander}, {Kaspi}, \& {Maoz}}]{peterson98}
{Peterson}, B.~M., {Wanders}, I., {Horne}, K., {et~al.} 1998, \pasp, 110, 660

\bibitem[{{Petrucci} {et~al.}(2023){Petrucci}, {Pi{\'e}tu}, {Behar}, {Clavel}, {Bianchi}, {Henri}, {Barnier}, {Chen}, {Ferreira}, {Malzac}, {Belmont}, {Corbel}, \& {Coriat}}]{petrucci23}
{Petrucci}, P.~O., {Pi{\'e}tu}, V., {Behar}, E., {et~al.} 2023, \aap, 678, L4

\bibitem[{{Piconcelli} {et~al.}(2004){Piconcelli}, {Jimenez-Bail{\'o}n}, {Guainazzi}, {Schartel}, {Rodr{\'\i}guez-Pascual}, \& {Santos-Lle{\'o}}}]{Piconcelli2004}
{Piconcelli}, E., {Jimenez-Bail{\'o}n}, E., {Guainazzi}, M., {et~al.} 2004, \mnras, 351, 161

\bibitem[{{Raginski} \& {Laor}(2016)}]{Raginski16}
{Raginski}, I. \& {Laor}, A. 2016, \mnras, 459, 2082

\bibitem[{{Ricci} {et~al.}(2023){Ricci}, {Chang}, {Kawamuro}, {Privon}, {Mushotzky}, {Trakhtenbrot}, {Laor}, {Koss}, {Smith}, {Gupta}, {Dimopoulos}, {Aalto}, \& {Ros}}]{ricci23}
{Ricci}, C., {Chang}, C.-S., {Kawamuro}, T., {et~al.} 2023, \apjl, 952, L28

\bibitem[{{Ricci} {et~al.}(2021){Ricci}, {Loewenstein}, {Kara}, {Remillard}, {Trakhtenbrot}, {Arcavi}, {Gendreau}, {Arzoumanian}, {Fabian}, {Li}, {Ho}, {MacLeod}, {Cackett}, {Altamirano}, {Gandhi}, {Kosec}, {Pasham}, {Steiner}, \& {Chan}}]{Ricci21}
{Ricci}, C., {Loewenstein}, M., {Kara}, E., {et~al.} 2021, \apjs, 255, 7

\bibitem[{{Ricci} {et~al.}(2017){Ricci}, {Trakhtenbrot}, {Koss}, {Ueda}, {Del Vecchio}, {Treister}, {Schawinski}, {Paltani}, {Oh}, {Lamperti}, {Berney}, {Gandhi}, {Ichikawa}, {Bauer}, {Ho}, {Asmus}, {Beckmann}, {Soldi}, {Balokovi{\'c}}, {Gehrels}, \& {Markwardt}}]{ricci17ApJS}
{Ricci}, C., {Trakhtenbrot}, B., {Koss}, M.~J., {et~al.} 2017, \apjs, 233, 17

\bibitem[{{Sabbatini} {et~al.}(2005){Sabbatini}, {Cavaliere}, {dall'Oglio}, {Davies}, {Martinis}, {Miriametro}, {Paladini}, {Pizzo}, {Russo}, \& {Valenziano}}]{HII}
{Sabbatini}, L., {Cavaliere}, F., {dall'Oglio}, G., {et~al.} 2005, \aap, 439, 595

\bibitem[{{Scargle}(1982)}]{scargle}
{Scargle}, J.~D. 1982, \apj, 263, 835

\bibitem[{{Shen} {et~al.}(2023){Shen}, {Grier}, {Horne}, {Stone}, {Li}, {Yang}, {Homayouni}, {Trump}, {Anderson}, {Brandt}, {Hall}, {Ho}, {Jiang}, {Petitjean}, {Schneider}, {Tao}, {Donnan}, {AlSayyad}, {Bershady}, {Blanton}, {Bizyaev}, {Bundy}, {Chen}, {Davis}, {Dawson}, {Fan}, {Greene}, {Groller}, {Guo}, {Ibarra-Medel}, {Keenan}, {Kollmeier}, {Lejoly}, {Li}, {de la Macorra}, {Moe}, {Nie}, {Rossi}, {Smith}, {Tee}, {Weijmans}, {Xu}, {Yue}, {Zhou}, {Zhou}, \& {Zou}}]{sdss-rm}
{Shen}, Y., {Grier}, C.~J., {Horne}, K., {et~al.} 2023, arXiv e-prints, arXiv:2305.01014

\bibitem[{{Smith} \& {Sartori}(2023)}]{pictor_a}
{Smith}, K.~L. \& {Sartori}, L.~F. 2023, \apj, 958, 188

\bibitem[{{Thomas} {et~al.}(2017){Thomas}, {Dopita}, {Shastri}, {Davies}, {Hampton}, {Kewley}, {Banfield}, {Groves}, {James}, {Jin}, {Juneau}, {Kharb}, {Sairam}, {Scharw{\"a}chter}, {Shalima}, {Sundar}, {Sutherland}, \& {Zaw}}]{Thomas2017}
{Thomas}, A.~D., {Dopita}, M.~A., {Shastri}, P., {et~al.} 2017, \apjs, 232, 11

\bibitem[{{Tombesi} {et~al.}(2012){Tombesi}, {Cappi}, {Reeves}, \& {Braito}}]{tombesi12}
{Tombesi}, F., {Cappi}, M., {Reeves}, J.~N., \& {Braito}, V. 2012, \mnras, 422, L1

\bibitem[{{Torrence} \& {Compo}(1998)}]{wavelet_prac}
{Torrence}, C. \& {Compo}, G.~P. 1998, Bulletin of the American Meteorological Society, 79, 61

\bibitem[{{Tortosa} {et~al.}(2023){Tortosa}, {Ricci}, {Shablovinskaia}, {Tombesi}, {Kawamuro}, {Kara}, {Mantovani}, {Balokovic}, {Chang}, {Gendreau}, {Koss}, {Liu}, {Loewenstein}, {Paltani}, {Privon}, \& {Trakhtenbrot}}]{tortosa23}
{Tortosa}, A., {Ricci}, C., {Shablovinskaia}, E., {et~al.} 2023, arXiv e-prints, arXiv:2312.00783

\bibitem[{{Unger} {et~al.}(1987){Unger}, {Lawrence}, {Wilson}, {Elvis}, \& {Wright}}]{Unger1987}
{Unger}, S.~W., {Lawrence}, A., {Wilson}, A.~S., {Elvis}, M., \& {Wright}, A.~E. 1987, \mnras, 228, 521

\bibitem[{{Vasudevan} \& {Fabian}(2009)}]{VF09}
{Vasudevan}, R.~V. \& {Fabian}, A.~C. 2009, \mnras, 392, 1124

\bibitem[{{Vaughan} {et~al.}(2003){Vaughan}, {Edelson}, {Warwick}, \& {Uttley}}]{vaughan03}
{Vaughan}, S., {Edelson}, R., {Warwick}, R.~S., \& {Uttley}, P. 2003, \mnras, 345, 1271

\bibitem[{{Wang} {et~al.}(2023){Wang}, {An}, {Zhang}, {Cheng}, {Ho}, {Kellermann}, \& {Baan}}]{wang23}
{Wang}, A., {An}, T., {Zhang}, Y., {et~al.} 2023, \mnras, 525, 6064

\bibitem[{{Wilson} {et~al.}(2013){Wilson}, {Rohlfs}, \& {H{\"u}ttemeister}}]{wilson_book}
{Wilson}, T.~L., {Rohlfs}, K., \& {H{\"u}ttemeister}, S. 2013, {Tools of Radio Astronomy}

\bibitem[{{Wright} {et~al.}(1999){Wright}, {Dickel}, {Koralesky}, \& {Rudnick}}]{casa2}
{Wright}, M., {Dickel}, J., {Koralesky}, B., \& {Rudnick}, L. 1999, \apj, 518, 284

\bibitem[{{Yu} {et~al.}(2020){Yu}, {Kochanek}, {Peterson}, {Zu}, {Brandt}, {Cackett}, {Fausnaugh}, \& {McHardy}}]{yu20}
{Yu}, Z., {Kochanek}, C.~S., {Peterson}, B.~M., {et~al.} 2020, \mnras, 491, 6045

\bibitem[{{Zu} {et~al.}(2016){Zu}, {Kochanek}, {Koz{\l}owski}, \& {Peterson}}]{zu16}
{Zu}, Y., {Kochanek}, C.~S., {Koz{\l}owski}, S., \& {Peterson}, B.~M. 2016, \apj, 819, 122

\end{thebibliography}
%

\begin{appendix}

\section{Data} \label{appendA}

\begin{table*}
\centering
\caption{ALMA data.}
\begin{tabular}{ccccc}
\hline
Start   Time (UTC) & End Time   (UTC) & beam size (arcsec) & $F$ (mJy) & $\sigma$ (mJy/beam) \\ \hline
8/28/2021 19:55    & 8/28/2021 20:27  & 0.0666 $\times$ 0.0576    & 3.61    & 0.0238         \\
8/29/2021 16:08    & 8/29/2021 16:40  & 0.0916 $\times$  0.0576    & 3.87    & 0.0229         \\
8/30/2021 21:23    & 8/30/2021 21:54  & 0.0674 $\times$  0.0627    & 5.10    & 0.0280         \\
8/31/2021 16:09    & 8/31/2021 16:39  & 0.0811 $\times$  0.0552    & 4.30    & 0.0266         \\
9/01/2021 20:05    & 9/01/2021 20:36  & 0.0604 $\times$  0.0504    & 3.51    & 0.0249         \\
9/02/2021 17:04    & 9/02/2021 17:34  & 0.0539 $\times$  0.0452    & 3.59    & 0.0244         \\
9/03/2021 21:18    & 9/03/2021 21:48  & 0.0513 $\times$  0.0441    & 4.00    & 0.0217         \\
9/04/2021 16:22    & 9/04/2021 16:52  & 0.0546 $\times$  0.0420    & 4.45    & 0.0229         \\
9/05/2021 19:57    & 9/05/2021 20:28  & 0.0457 $\times$  0.0438    & 6.38    & 0.0216         \\
9/06/2021 19:14    & 9/06/2021 19:46  & 0.0649 $\times$  0.0422    & 9.23    & 0.0277         \\ \hline
\end{tabular}
\end{table*}

\begin{table*}[!ht]
\centering
\caption{X-ray data in 2--10~keV range (except \textit{NICER} epoch marked with the asterisk, where the data is given for 0.7--7\,keV range).}
\begin{tabular}{lccccc}
\hline
Obs. ID     & Start Time    & Telescope  & Exposure & Flux  & {Flux range } \\ 
     & UTC    & & sec & $10^{-11}$ \flux & {68\% confidence} \\\hline
00014460001 & 2021-08-07 02:22:34 & \textit{Swift}      & 5571           & 10.8            & (1.04e$-$10 -- 1.11e$-$10)                                  \\
00014460002 & 2021-08-08 00:34:34 & \textit{Swift }     & 5631           & 8.12            & (7.82e$-$11 -- 8.39e$-$11)                                  \\
00014460003 & 2021-08-09 05:14:52 & \textit{Swift}      & 7729           & 8.70            & (8.45e$-$11 -- 8.93e$-$11)                                  \\
0862090101  & 2021-08-10 16:50:25 & \textit{XMM-Newton} & 10290          & 11.0            & 0.24e$-$11                                               \\
00089304001 & 2021-08-10 20:55:34 & \textit{Swift}      & 1573           & 11.6            & (1.08e$-$10 -- 1.24e$-$10)                                  \\
0862090201  & 2021-08-11 18:36:37 & \textit{XMM-Newton} & 10540          & 8.84            & 0.38e$-$11                                               \\
0862090301  & 2021-08-12 16:45:45 & \textit{XMM-Newton} & 13070          & 7.79            & 0.53e$-$11                                               \\
0862090501  & 2021-08-14 16:45:15 & \textit{XMM-Newton} & 14440          & 9.09            & 0.29e$-$11                                               \\
0862090601  & 2021-08-15 18:02:00 & \textit{XMM-Newton} & 7455           & 9.18            & 0.42e$-$11                                               \\
0862090701  & 2021-08-16 16:42:33 & \textit{XMM-Newton} & 18750          & 8.64            & 0.36e$-$11                                               \\
00089304002 & 2021-08-16 22:02:35 & \textit{Swift }     & 1808           & 8.94           & (8.44e$-$11 -- 9.39e$-$11)                                  \\
0862090801  & 2021-08-17 17:41:23 & \textit{XMM-Newton} & 12360          & 8.76            & 0.29e$-$11                                               \\
0862090901  & 2021-08-18 16:35:14 & \textit{XMM-Newton} & 18030          & 8.49            & 0.27e$-$11                                               \\
0862091001  & 2021-08-19 17:08:37 & \textit{XMM-Newton} & 13620          & 8.90            & 0.46e$-$11                                               \\
00014460004 & 2021-08-20 02:45:36 & \textit{Swift}      & 7399           & 9.57            & (9.28e$-$11 -- 9.87e$-$11)                                  \\
00014460005 & 2021-08-21 05:23:34 & \textit{Swift }     & 7062           & 10.6            & (1.03e$-$10 -- 1.09e$-$10)                                  \\
00014460006 & 2021-08-22 00:38:36 & \textit{Swift }     & 8666           & 9.71            & (9.43e$-$11 -- 9.98e$-$11)                                  \\
00014460007 & 2021-08-23 00:24:34 & \textit{Swift }     & 7684           & 11.3            & (1.09e$-$10 -- 1.16e$-$10)                                  \\
00014460008 & 2021-08-24 01:52:36 & \textit{Swift}      & 3389           & 12.0            & (1.15e$-$10 -- 1.24e$-$10)                                  \\
00014460009 & 2021-08-25 03:29:34 & \textit{Swift}      & 7657           & 10.7            & (1.04e$-$10 -- 1.11e$-$10)                                  \\
3599010101  & 2021-08-25 15:57:00 & \textit{NICER}      & 777            & 9.87            & (9.69e$-$11 -- 1.00e$-$10)                                  \\
00014460010 & 2021-08-26 00:03:34 & \textit{Swift }     & 10040          & 8.32            & (8.09e$-$11 -- 8.53e$-$11)                                  \\
3599010102  & 2021-08-26 04:21:00 & \textit{NICER}      & 2497           & 8.18            & (8.09e$-$11 -- 8.26e$-$11)                                  \\
3599010103*  & 2021-08-27 19:05:23 & \textit{NICER}      & 1361           & 10.6            & (1.04e$-$10 -- 1.08e$-$10)                                 \\
3599010104  & 2021-08-28 10:35:23 & \textit{NICER}      & 2817           & 10.9            & (1.08e$-$10 -- 1.10e$-$10)                                  \\
3599010105  & 2021-08-29 20:41:25 & \textit{NICER}      & 1651           & 10.7            & (1.06e$-$10 -- 1.08e$-$10)                                  \\
3599010106  & 2021-08-30 19:56:04 & \textit{NICER}      & 1524           & 9.06            & (8.93e$-$11 -- 9.18e$-$11)                                  \\
3599010107  & 2021-08-31 19:11:05 & \textit{NICER}      & 1387           & 9.72            & (9.59e$-$11 -- 9.84e$-$11)                                  \\
3599010108  & 2021-09-01 19:59:29 & \textit{NICER}      & 1541           & 11.8            & (1.17e$-$10 -- 1.20e$-$10)                                  \\
3599010109  & 2021-09-02 19:14:29 & \textit{NICER}      & 1411           & 11.7            & (1.16e$-$10 -- 1.18e$-$10)                                  \\
3599010110  & 2021-09-03 20:02:30 & \textit{ NICER}      & 1513           & 9.68           & (9.56e$-$11 -- 9.81e$-$11)                                  \\
3599010111  & 2021-09-04 19:18:10 & \textit{NICER}      & 1401           & 11.7            & (1.16e$-$10 -- 1.18e$-$10)                                  \\
3599010112  & 2021-09-05 20:02:00 & \textit{NICER}      & 1618           & 14.6e           & (1.45e$-$10 -- 1.48e$-$10)                                  \\
3599010113  & 2021-09-06 10:01:42 & \textit{NICER}      & 854            & 14.9            & (1.47e$-$10 -- 1.52e$-$10)                                  \\
3599010114  & 2021-09-06 23:58:41 & \textit{NICER}      & 1983           & 15.2           & (1.50e$-$10 -- 1.53e$-$10)                                  \\
3599010115  & 2021-09-08 00:47:16 & \textit{NICER}      & 2174           & 13.0            & (1.29e$-$10 -- 1.32e$-$10)                                  \\
3599010116  & 2021-09-09 00:06:16 & \textit{NICER}      & 1980           & 12.2            & (1.21e$-$10 -- 1.23e$-$10)                                  \\
3599010117  & 2021-09-10 01:11:20 & \textit{NICER}      & 1219           & 11.4            & (1.12e$-$10 -- 1.15e$-$10)                                  \\
3599010118  & 2021-09-11 00:04:36 & \textit{NICER}      & 2094           & 13.4            & (1.32e$-$10 -- 1.35e$-$10)                                  \\
3599010119  & 2021-09-12 00:52:56 & \textit{NICER}      & 1839           & 12.1            & (1.20e$-$10 -- 1.23e$-$10)                                  \\
3599010120  & 2021-09-13 00:08:37 & \textit{NICER}      & 1229           & 12.0            & (1.18e$-$10 -- 1.21e$-$10)                                  \\
3599010121  & 2021-09-14 19:56:00 & \textit{NICER}      & 403            & 11.3            & (1.11e$-$10 -- 1.16e$-$10)                                  \\ \hline
\end{tabular}

\end{table*}

\end{appendix}
\end{document}